\documentclass[amsmath,amssymb,aps,prb,superscriptaddress,twocolumn]{revtex4-1}

\usepackage{hyperref}
\usepackage{latexsym,amsbsy,amsfonts,graphicx}
\usepackage{dcolumn}
\usepackage{bm}
\usepackage[utf8]{inputenc}
\usepackage[english]{babel}
\usepackage{color}
\usepackage{verbatim}

\newcommand{\abs}[1]{\ensuremath{\lvert#1\rvert}}
\newcommand{\qv}{\mathbf{q}}
\newcommand{\kv}{\mathbf{k}}
\newcommand{\av}[1]{\ensuremath{\left\langle #1 \right\rangle}}

\DeclareMathOperator{\Imag}{Im}

\usepackage{tikz}
\usetikzlibrary{calc}
\usetikzlibrary{decorations.pathmorphing}
\usetikzlibrary{decorations.pathreplacing}
\usetikzlibrary{positioning}
\usetikzlibrary{patterns}

\definecolor{erik_diagram}{RGB}{235, 100, 0}

\begin{document}

\title{Self--consistent Dual Boson approach\\  to single-particle and collective excitations in correlated systems}

\author{E. A. Stepanov}
\affiliation{Radboud University, Institute for Molecules and Materials, 6525AJ Nijmegen, The Netherlands}

\author{E. G. C. P. van Loon}
\affiliation{Radboud University, Institute for Molecules and Materials, 6525AJ Nijmegen, The Netherlands}

\author{A. A. Katanin}
\affiliation{Institute of Metal Physics, 620990 Ekaterinburg, Russia}
\affiliation{Ural Federal University, Department of Theoretical Physics and Applied Mathematics, 620002, Ekaterinburg, Russia}

\author{A. I. Lichtenstein}
\affiliation{Institute of Theoretical Physics, University of Hamburg, 20355 Hamburg, Germany}
\affiliation{Ural Federal University, Department of Theoretical Physics and Applied Mathematics, 620002, Ekaterinburg, Russia}

\author{M. I. Katsnelson}
\affiliation{Radboud University, Institute for Molecules and Materials, 6525AJ Nijmegen, The Netherlands}
\affiliation{Ural Federal University, Department of Theoretical Physics and Applied Mathematics, 620002, Ekaterinburg, Russia}

\author{A. N. Rubtsov}
\affiliation{Russian Quantum Center, 143025 Skolkovo,  Russia}
\affiliation{Department of Physics, M.V. Lomonosov Moscow State University, 119991 Moscow, Russia}

\date{\today}

\begin{abstract}
We propose an efficient dual boson scheme, which extends the DMFT paradigm to collective excitations in correlated systems. The theory is fully self--consistent both on the one-- and on the two--particle level, thus describing the formation of collective modes as well as the renormalization of electronic and bosonic spectra on equal footing. The method employs an effective impurity model  comprising both fermionic and bosonic hybridization functions. Only single-- and two--electron Green's functions of the reference problem enter the theory, due to the optimal choice of the self--consistency condition for the effective bosonic bath. We show that the theory is naturally described by a  
dual Luttinger--Ward functional and obeys the relevant conservation laws.
\end{abstract}


\maketitle


\section{Introduction}

Strongly correlated electron systems remain one of the most interesting subjects in modern condensed matter physics. 
It is hard to treat such systems analytically due to the large local and, more importantly, nonlocal electron--electron correlations, and therefore the development of appropriate computational methods is important.
Nonlocal correlation effects are very important for studying charge-ordering~\cite{Tosatti74,Hansmann13,Huang14,vanLoon14-2} and Wigner--Mott~\cite{Mott74,Imada98,Walz02} transitions, plasmon~\cite{vanLoon14,Hafermann14-2} and magnon~\cite{Moriya1985,Irkhin85,Secchi13} modes, antiferromagnetic fluctuations~\cite{Rubtsov09,Rohringer11,Schafer15} and other interesting features of such systems. 
These phenomena are realizable in adatoms on semiconducting surfaces~\cite{Tosatti74,Hansmann13,Huang14}, or in systems of cold atoms~\cite{Bloch12,Lewenstein12,vanLoon15-2} and in graphene~\cite{Wehling11}.

Dynamical mean--field theory (DMFT)~\cite{Metzner89,Georges96} has become the standard approximation for strongly correlated fermionic systems.
In this theory, all local correlations are treated via an auxiliary impurity problem, i.e., the electrons on a site are influenced by 
an effective local electronic bath formed by the other electrons.
This effective description becomes exact in the limit of infinite dimension~\cite{Metzner89,Georges96} and captures the formation of Hubbard bands~\cite{Hubbard63,Hubbard64} and the Mott transition~\cite{Mott74,Imada98}.
In finite dimension, DMFT is an approximation that neglects nonlocal correlation effects. In particular, the impact of collective modes 
on the impurity problem is neglected.

There have been many attempts to go beyond DMFT and to incorporate these effects.  
The Dual Fermion (DF) approach has been developed to take nonlocal fermion correlations into account~\cite{Rubtsov08}.
This approach is exact in the two important limits of large and small local interaction.
Similar efforts have been made in the D$\Gamma$A~\cite{Toschi07} and 1PI~\cite{Rohringer13} approaches and in the recently proposed DMF$^2$RG method~\cite{Taranto14,Wentzell15}.
Nonlocal fermionic correlations are crucial for a description of pseudogap formation~\cite{Rubtsov08}, critical exponents for magnetic phase transitions~\cite{Rohringer11,Antipov14} and formation of flat bands near van Hove singularity~\cite{Yudin14}.
However, these methods cannot describe bosonic degrees of freedom and their influence on the auxiliary local model.

Extended dynamical mean--field theory (EDMFT)~\cite{Si96,Smith00,Chitra00,Chitra01,Sun02} was introduced to include bosonic degrees of freedom into DMFT, with the main focus on nonlocal density--density interactions.
In this approach all fermionic and bosonic correlations are treated on the effective impurity level, and therefore are local.
However, it was realized that nonlocal corrections beyond EDMFT are necessary for a correct description of strongly nonlocal effects, such as plasmons.
Even though  DMFT and EDMFT are similar in spirit, the reduction of correlation effects to their local part works much better for fermions than for bosons.

Thus, EDMFT served as a starting point for theories that include further spatial correlations.
The first example of such a theory is EDMFT+GW~\cite{Sun02}.
EDMFT+GW and, more recently, TRILEX~\cite{Ayral15} diagrammatically treat some nonlocal effects beyond EDMFT.
However, conservation laws are not automatically fulfilled in such approaches~\cite{Rubtsov12,Hafermann14-2}.

In general, studying the collective excitations in strongly correlated systems is challenging. 
Historically, two avenues have been explored. 
One can start from the bare Green's functions, as in the random phase approximation (RPA)~\cite{Lindhard54,Ehrenreich59,Pines66,Platzman73}, however the bare Green's functions do not contain any information about the spectral weight transfer to the Hubbard bands and correlation effects.
A theory defined in terms of renormalized Green's functions, on the other hand, contains all the information about the Hubbard bands. However, a consistent~\cite{Baym61,Baym62} description of the collective excitations requires not only the renormalized Green's functions but also the renormalized vertices and these are numerically very challenging to handle. 

Recently, the Dual Boson (DB) approach~\cite{Rubtsov12} was developed to address these issues. 
It applies a transformation to new degrees of freedom that contain the information about the Hubbard bands and correlation effects already in their \emph{bare} Green's functions. 
In this way, the DB approach for strongly correlated systems fulfills conservation laws~\cite{Hafermann14-2}, necessary for a correct description of plasmons~\cite{vanLoon14}, while remaining computationally tractable.
The DB method is a diagrammatic extension of EDMFT that can be applied to correlated lattice fermion models with local and nonlocal interaction.
It allows us to include spatial fermionic and bosonic correlations beyond EDMFT by introducing new dual variables. These dual variables are 
introduced via fermionic and bosonic hybridization functions $\Delta_{\nu}$ and $\Lambda_{\omega}$ that act as the effective mean--fields acting on an auxiliary single--site impurity.
Then, the impurity model serves as a starting point for a perturbative expansion.
By choosing the hybridization functions in an optimal way, this perturbative expansion can be simplified.
In this work, we will study how the bosonic hybridization function $\Lambda_\omega$ should be determined to optimally treat the feedback of collective modes onto the impurity.

As in DF, the choice of the hybridization function has implications for the dual perturbation theory. 
We will show that the correct choice of $\Lambda_\omega$ removes all local two--particle processes from the dual perturbation theory.
Furthermore, we show that the self--consistent determination of $\Lambda_\omega$ gives the correct results in some important limits and that it satisfies the 
charge conservation law.
Finally, we study the physical impact of the choice of $\Lambda_\omega$, showing that the nonlocal charge fluctuations make the system more insulating.

The paper is organized as follows: We start by giving a short description of the Dual Boson formalism in section~\ref{sec:db}. Then, in section~\ref{sec:sc} we turn to the self--consistency conditions in DB and related methods. In section~\ref{sec:luttingerward}, we discuss the dual Luttinger--Ward functional and the generation of self--energy diagrams.
The impact of the self--consistency on the self--energy diagrams is given in section~\ref{HOV}. In section~\ref{sec:charge}, we show that the self--consistent DB method satisfies charge conservation requirements. 
Analytical and numerical results regarding the self--consistency condition are given in section~\ref{sec:application}. 

\section{Dual Boson formalism}
\label{sec:db}

The DB~\cite{Rubtsov12} approach to strongly correlated systems relies on a separation of local and nonlocal correlation effects. 
In this section we give a general description of the approach, and in the next section we will elaborate on the role of the self--consistency 
condition in DB. More details on the derivation and the technical implementation can be found in Refs.~\onlinecite{Rubtsov12} and \onlinecite{vanLoon14-2}.

We consider the extended Hubbard model in the Matsubara formalism.
To be specific, we restrict ourselves in this work to the half--filled extended Hubbard model on a square lattice, with the action
\begin{align}
S[c^{*},c]=&-\sum_{j\nu\sigma} c^{*}_{j\nu\sigma}[i\nu+\mu]c^{\phantom{*}}_{j\nu\sigma} + \frac{1}{2}U\sum_{j\omega}n_{j\omega}n_{j,-\omega}\notag\\
&+ \sum_{\av{jl}\nu\sigma}t_{jl}c^{*}_{l\nu\sigma}c^{\phantom{*}}_{j\nu\sigma} + \frac{1}{2}\sum_{\av{jl}\omega}V_{jl}n_{l\omega}n_{j,-\omega}.
\label{eq:s:lattice}
\end{align}
Here, $c^{*}_{j\nu\sigma}$ ($c_{j\nu\sigma}$) are Grassmann variables corresponding to creation (annihilation) of an electron on 
site $j$ with spin $\sigma$ and fermionic Matsubara frequency $\nu$. $n_j$ counts the number of electrons on site $j$, and $\omega$ is a bosonic 
Matsubara frequency. A normalization by $\beta$ is implied in the sum over frequencies. The chemical potential $\mu$ is chosen in such a way 
that the average number of electrons per site is one (half--filling). $U$ and $V$ are the on--site and nearest--neighbor interaction and $t$ is 
the hopping integral between neighboring sites. We use $t=1$ as the unit of energy.

The action \eqref{eq:s:lattice} is split into a set of single--site impurity problems and the remaining part $S_{rem}$,
\begin{align}
\label{eq:s:imp}
S_{\text{imp}}[c^{*},c]=&-\sum_{\nu\sigma} c^{*}_{\nu\sigma}[i\nu+\mu-\Delta_{\nu\sigma}]c^{\phantom{*}}_{\nu\sigma}\notag\\
&+ \frac{1}{2}\sum_{\omega}n_{\omega}[U+\Lambda_{\omega}] n_{-\omega}\\
S =& \sum_{j} S^{(j)}_{\text{imp}}[c^{*},c] + S_{\text{rem}}.
\end{align}
Here, the hybridization functions $\Delta_{\nu\sigma}$ and $\Lambda_{\omega}$ were introduced. For the moment, these are arbitrary, and 
we will study the choice of the hybridization functions in this work.

\begin{figure}[]
\includegraphics[width=0.6\linewidth]{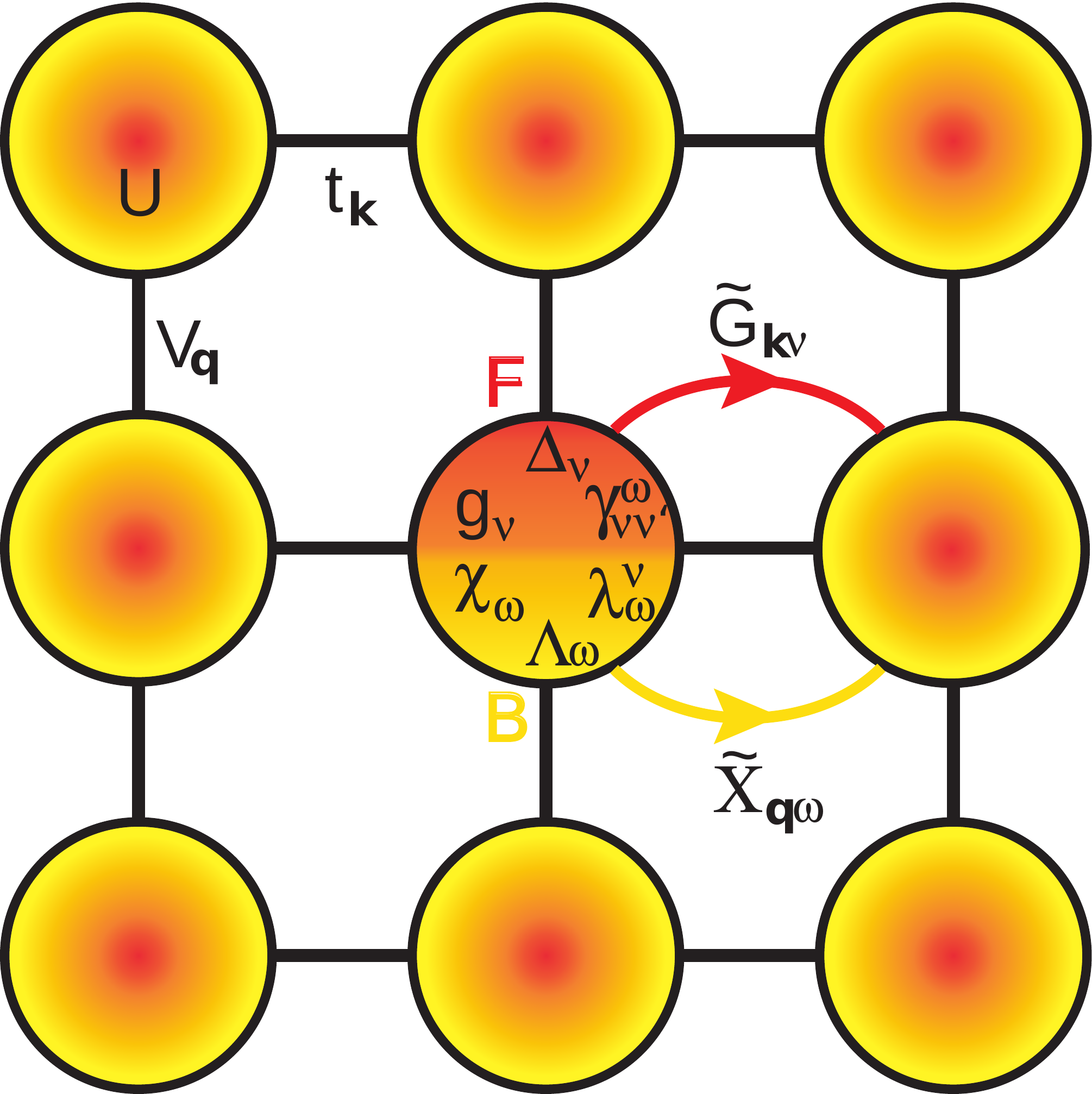}
\caption{(Color online) Sketch of the DB formalism. The original action with parameters $U$, $V_\qv$ and $t_\kv$ is replaced by an auxiliary impurity 
problem with fields $\Delta_\nu$, $\Lambda_\omega$. The expectation values of this impurity model ($g_\nu$, $\chi_\omega$, $\gamma_{\nu\nu'\omega}$ 
and $\lambda_{\nu\omega}$) enter a \emph{dual} theory in terms of $\tilde{G}$ and $\tilde{X}$.}
\label{DB_basic}
\end{figure}

The DB formalism proceeds by introducing new \emph{dual} degrees of freedom $f^*$, $f$, $\phi$ via a Hubbard-Stratonovich 
transformation of $S_{rem}$ and integrating out the original degrees of freedom $c^*$ and $c$. In this way, the original action \eqref{eq:s:lattice} 
is subdivided into two separate problems: the impurity action \eqref{eq:s:imp} and a \emph{dual} action
\begin{align}
\tilde{S}[f^{*},f;\phi] =&
- \sum_{\kv\nu\sigma}f^{*}_{\kv\nu\sigma} (\tilde{G}^{(0)}_{\kv\nu\sigma})^{-1} f^{\phantom{*}}_{\kv\nu\sigma} \notag\\
&- \frac{1}{2} \sum_{\qv\omega} \phi_{\qv\omega}(\tilde{X}^{0}_{\qv\omega})^{-1}\phi_{\qv\omega}\notag\\
&+ \tilde{V}[f^{*},f,\phi].
\label{eq:s:dual}
\end{align}
This dual action has bare propagators
\begin{align}
\tilde{G}^{(0)}_{\kv\nu} &= [g^{-1}_{\nu}+\Delta_{\nu}-\varepsilon_{k}]^{-1} - g_{\nu},\label{eq:barefermionpropagator}
\\
\tilde{X}^{0}_{\qv\omega} &= [\chi^{-1}_{\omega}+\Lambda_{\omega}-V_{q}]^{-1} - \chi_{\omega}.
\label{eq:barebosonpropagator}
\end{align}
Here $\varepsilon_{k}$ and $V_{q}$ are the Fourier transform of $t_{jl}$ and $V_{jl}$, $g$ and $\chi$ are the impurity Green's function and charge susceptibility respectively
\begin{align}
g_{\nu} =& -\av{c_{\nu}c^{*}_{\nu}}_\text{imp},\\
\chi_{\omega} =& -\av{\rho_{\omega}\rho_{-\omega}}_\text{imp},
\end{align}
where $\rho = n_{\omega} - \av{n}\delta_{\omega}$ and $\av{\ldots}_\text{imp}$ denotes the impurity average with respect to the action 
\eqref{eq:s:imp}. 
The single--frequency susceptibility $\chi=-K_{1=1',2=2'}$ can be obtained from the usual expression of the two--particle correlation function $K_{122'1'} = \av{c^{\phantom{*}}_1 c^{\phantom{*}}_2 c^*_{2'} c^*_{1'}}$ (hereafter, we use combined frequency--spin subscripts,   e.g. $1\equiv \nu_1, \sigma_1$). 

The main philosophy of the DB method is that the impurity action can be solved numerically exactly, so we should try to put as much of the physics as possible into the impurity model.
The exact solution of the impurity model can be found using a continuous--time quantum Monte Carlo~\cite{Rubtsov05,Werner06} solver.
Only weaker correlations remain in the dual perturbation theory, which 
yelds the fully renormalized fermionic and bosonic dual proparators $\tilde{G}_{\kv\nu} = -\av{f^{\phantom{*}}_{\kv\nu}f^{*}_{\kv\nu}}$ and $\tilde{X}_{\qv\omega} = -\av{\phi_{\qv\omega}\phi_{-\qv, -\omega}}$.
The scheme is illustrated graphically in Fig.~\ref{DB_basic}.

\begin{figure}
\includegraphics{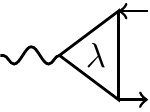}
 \,\,\,
\includegraphics{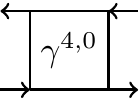}
 \,
\includegraphics{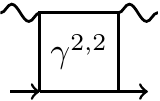}

 \vspace{0.5cm}
\includegraphics{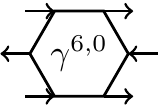}
\includegraphics{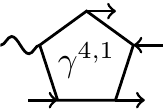}
 \caption{Interaction vertices in the dual perturbation theory. }
 \label{fig:vertices}
\end{figure}

The dual interaction $\tilde{V}$ contains interaction terms of arbitrary order in $f^*$, $f$ and $\phi$. 
All vertices up to three--particle level with at least two fermion lines are shown in Fig.~\ref{fig:vertices}. 
We use the notation $\gamma^{n,m}$ for the vertex with $n$ fermion and $m$ boson lines. Usually, the interactions are restricted to the 
two--particle level, where there is a fermion--fermion interaction $\gamma=\gamma^{4,0}$ and a fermion--boson interaction $\lambda=\gamma^{2,1}$
\begin{align}
\gamma^{4,0}_{1,2,3,4} &= (g_1g_2g_3g_4)^{-1}\big[\av{c_1c_2c^{*}_3c^{*}_4} - \notag \\ 
&\,\,\,\,\,\,\,\,\,\,\,\,\,\,\,\,\,\,\,\,\,\,\,\,\,\,\,\,\,\,\,\,\,\,\,\,
\av{c_1c^{*}_4}\av{c_2c^{*}_3} + \av{c_1c^{*}_3}\av{c_2c^{*}_4}],\\
\gamma^{2,1}_{1,4;2} &= -(g_1g_4 \chi_2)^{-1}\big[\av{c_1c^{*}_4n_{2}} - \av{c_1c^{*}_4} \av{n_2}] \notag \\
&= (g_1g_4 \chi_2)^{-1}\sum_{3}\big[\av{c_1c_{2+3}c^{*}_3c^{*}_4} - \av{c_1c^{*}_4} \av{c_{2+3}c^{*}_3}].
\end{align}
These interaction vertices are related by~\cite{Rubtsov12,vanLoon14-2}
\begin{align}
 \gamma^{2,1}_{1,4;2}= \chi_2^{-1}\sum_{3}\big[\gamma^{4,0}_{1,2+3,3,4}g_{2+3}g_3-\delta_{1,3}\delta_{2+3,4}\big].
 \label{eq:lambda}
\end{align}
In particular, the fermion--boson vertex is non--zero even in the non--interacting system ($\gamma^{4,0}=0$). 
A further sum--rule for $\gamma^{2,1}$ is given in Appendix~\ref{app:sumrule}. 
Exact relations for the three--particle vertices are given in Appendix~\ref{app:vertices}. 
We will come back to the issue of restricting dual interactions to the two--particle level later.

To calculate the susceptibility $X_{\qv\omega}=-\av{nn}_{\qv\omega}$ of the original degrees of freedom, one starts by calculating the dual 
polarization (self energy of the dual bosonic propagator) $\tilde{\Pi}_{\qv\omega}=(\tilde{X}^{0})^{-1}-\tilde{X}^{-1}$.
This is done diagrammatically using the expansion of the dual action.
Then, the lattice susceptibility can be obtained from the dual polarization as
\begin{align}
X_{\qv\omega}^{-1} = \left[\chi_{\omega} + \chi_{\omega}\,\tilde{\Pi}_{\qv\omega}\,\chi_{\omega}\right]^{-1}+\Lambda_{\omega}-V_{\qv}.
\label{eq:xfrompid}
\end{align}

Similar equations hold for the relations between dual and lattice fermions. In particular, the Green's function is given by
 \begin{align}
 G_{\kv\nu}^{-1} =& \left[g_{\nu} + g_{\nu}\,\tilde{\Sigma}_{\kv\nu}\,g_{\nu}\right]^{-1}+\Delta_{\nu}-\varepsilon_{\kv},
\label{eq:gfromsigma}
 \end{align}
where $\tilde{\Sigma}$ is the dual fermionic self energy.

\section{Self--consistency conditions and bosonic superline}
\label{sec:sc}

The dual interaction functional $\tilde{V}$ can be represented by an infinite expansion over the vertex functions of the impurity problem.
Although the series is formally defined for arbitrary hybridization functions $\Delta_{\nu}$ and $\Lambda_{\omega}$, the specific choice
affects the convergence of particular diagrammatic approximations. Physically, $\Delta_{\nu}$ and $\Lambda_{\omega}$ describe 
the ``effective field'' generated by electrons on the other sites, felt by electrons in the impurity problem.
The freedom to choose them allows us include most of the correlation effects into the impurity, simplifying the dual action.
To do this, the hybridization functions are determined self--consistently according to some self--consistency condition. There are 
several options for the self--consistency condition, and here we will discuss them.

DMFT is restricted to systems with $V_\qv=0$ and it does not use a retarded interaction, $\Lambda_\omega=0$. 
It assumes that the self--energy is local, $\tilde{\Sigma}_{\kv\nu}=0$. 
The self--consistency condition on $\Delta_\nu$ is
\begin{align}
 \sum_{\kv} G_{\kv\nu} = g_\nu, \label{eq:sc:dmft}
\end{align}
or, equivalently~\cite{Rubtsov08} in the DMFT approximation,
\begin{align}
 \sum_{\kv} \tilde{G}_{\kv\nu} = 0. \label{eq:sc:df}
\end{align}
In both expressions, $\sum_{\kv}$ denotes a momentum average.

The DF extension of DMFT shines light on the meaning of the self--consistency condition~\cite{Rubtsov08}.
The self--energy in DMFT is the zeroth--order of the dual perturbation theory, and the condition \eqref{eq:sc:df} ensures that the 
first--order (Hartree) contribution to the self--energy is zero.\footnote{Higher order diagrams with a local dual Green's function connecting local vertices are also zero.}
In this way, DMFT and DF only differ in higher orders of the dual perturbation theory. In DF, the condition \eqref{eq:sc:dmft} is not 
equivalent to \eqref{eq:sc:df} due to nonlocal parts of the self--energy,
\begin{align}
\Sigma_{\kv\nu\sigma} = \Sigma^{\text{imp}}_{\nu\sigma} +
\frac{\tilde{\Sigma}_{k\nu\sigma}}{1+\tilde{\Sigma}_{k\nu\sigma}g_{\nu\sigma}},
\label{eq:self_energy_add}
\end{align}
and Eq.~\eqref{eq:sc:dmft} does not make the first--order self--energy diagram vanish. This is a strong motivation to use Eq.~\eqref{eq:sc:df} in DF. 

EDMFT, on the other hand, adds a retarded interaction $\Lambda_\omega$ to account for impurity screening by the nonlocal interaction $V_\qv$. 
This retarded interaction is determined by a self--consistency condition similar to that for $\Delta_{\nu}$, namely
\begin{align}
 \sum_{\qv} X_{\qv\omega} = \chi_\omega. \label{eq:sc:edmft}
\end{align}
Similar to the fermionic hybridization, within EDMFT there is an equivalent self--consistency condition (see~\cite{vanLoon14-2} or Appendix~\ref{equality})
\begin{align}
 \sum_{\qv} \tilde{X}_{\qv\omega} = 0. \label{eq:sc:db}
\end{align}
Finally, the DB approach also takes into account the
nonlocal $\tilde{\Pi}_{\qv\omega}$ and a choice between \eqref{eq:sc:db} and \eqref{eq:sc:edmft} needs to be made.

To study the meaning of the self--consistency condition \eqref{eq:sc:edmft}, we use an exact expression for the lattice susceptibility in terms of dual variables (see Appendix~\ref{app:DX})
\begin{align}
X = (1+\chi\tilde{\Pi})\big[\tilde{X}(1+\tilde{\Pi}\chi) + \chi\big] = \chi + \tilde{S},
\label{scline}
\end{align}
where we define the bosonic ``superline'' and propose the
new self--consistency condition as
\begin{align}
\sum_{\qv}&\tilde{S}_{\qv\omega} = 0, \notag \\
&\tilde{S} = \tilde{X} + \chi\tilde{\Pi}\chi + \chi\tilde{\Pi}\tilde{X} + \tilde{X}\tilde{\Pi}\chi + \chi\tilde{\Pi}\tilde{X}\tilde{\Pi}\chi.
\label{eq:superline}
\end{align}

According to Eq.~\eqref{scline}, the bosonic ``superline'' is exactly the difference between the lattice and impurity susceptibilities. 
It contains both types of two--particle processes, the fermionic ladder and the bosonic fluctuations, and combinations of these two. 
Using Eq.~\eqref{scline}, the ``lattice'' self--consistency condition \eqref{eq:sc:edmft} is exactly the requirement that the local part of the ``superline'' is zero (see Eq.~\eqref{eq:superline}). 
On the other hand, the ``dual'' self--consistency condition \eqref{eq:sc:db} can be understood as the requirement that the local part of only dual bosonic fluctuations is equal to zero. 
The former self--consistency condition treats collective fermionic and bosonic fluctuations on equal footing.

Until now, most\footnote{The self--consistency condition \eqref{eq:sc:db} has been used, see the black crosses in Fig. 21 of 
Ref.~\onlinecite{vanLoon14-2}. The effect compared to single--shot calculations was small.} numerical results obtained with DB~\cite{Hafermann14-2,vanLoon14,vanLoon14-2} 
started with EDMFT ($\tilde{\Pi}_{\qv\omega}=0)$ self--consistency and then did only a single--shot of DB, so the question of self--consistency was not relevant. 
These results showed that the susceptibility in DB satisfies charge conservation. 
This is a useful property, crucial for studying long--wavelength collective excitations, and we make sure that our self--consistent calculations will also satisfy this requirement. This issue will be studied in Sec. \ref{sec:charge}.

\section{Dual Luttinger--Ward functional}
\label{sec:luttingerward}

Let us now discuss a functional description of DB, containing both collective fermionic and bosonic fluctuations.
Before presenting a Baym--Kadanoff functional derivation of the theory~\cite{Baym61,Baym62}, we need to analyze how the 
conserving properties of the original and dual theory are related. We will show that the original and dual theories are simultaneously $\Phi$--derivable~\cite{Baym62}, so that the self--energy function fulfill variational equations $\delta \Phi=\Sigma \delta G$ and $\delta \tilde \Phi=\tilde \Sigma \delta \tilde G$ with certain functionals $\Phi, \tilde \Phi$. This is equivalent to the statement that $\delta \Phi- \delta \tilde \Phi$ is a full differential.

Using the notation $R=1+g \tilde \Sigma$, the exact relations \eqref{eq:gfromsigma} and \eqref{eq:self_energy_add} can be expressed in a compact form,
\begin{equation}
G=R g + R^2 \tilde G,
\label{eq:GGtilde}
\end{equation}
which is the equivalent of Eq.~\eqref{eq:superline} for fermions,
and 
\begin{equation}
\Sigma=i \nu -\Delta - (R g)^{-1}.
\end{equation}
An important property of this representation is that the dispersion law $\epsilon_k$ does not explicitly enter in the formulas. Therefore only renormalized quantities appear in the variational procedure,  in the spirit of the Baym philosophy. We obtain
\begin{equation}
\delta \Phi- \delta \tilde \Phi=(i \nu -\Delta) \delta G - (R g)^{-1} \delta  G - g ^{-1} (R-1) \delta \tilde G.
\end{equation}
We consider the variational procedure at fixed hybridization, so that variances of $\Delta$ and $g$ do not appear. Concequently, using the relation \eqref{eq:GGtilde}, the calculation gives  
\begin{equation}
\delta \Phi- \delta \tilde \Phi=(i \nu -\Delta) \delta G - (R^{-1}   + 2 g^{-1} \tilde G) \delta R  - g ^{-1} (2 R-1) \delta \tilde G,
\end{equation}
and we can get
\begin{equation}
\Phi- \tilde \Phi= (i \nu -\Delta) G +g^{-1} \tilde G -\ln R- 2 g^{-1} R \tilde G.
\label{eq:variation}
\end{equation}
This result is very important, because it shows the general possibility to construct a conserving theory from the dual theory. The obtained expression also establish an explicit relation between the original and dual functionals.

Now we can introduce the dual functional for our theory to show how to generate self--energy and polarization diagrams. 
The dual functional formulation does not lead to a straightforward proof of the Baym--Kadanoff conservation laws~\cite{Baym61,Baym62} for the original lattice problem. Charge conservation will be discussed in Sec.~\ref{sec:charge}.

\begin{figure}[ht]
\includegraphics[width=0.95\linewidth]{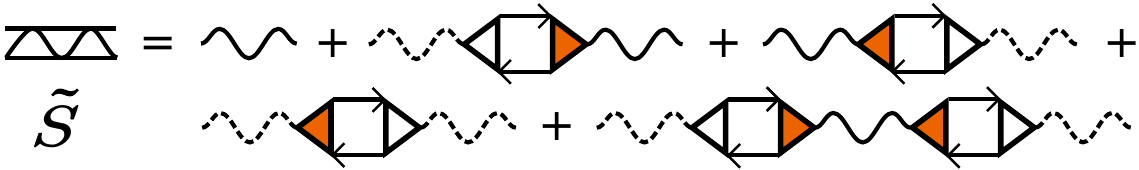}
\caption{(Color online) Diagrammatic representation of Eq.~\eqref{eq:superline}. The dual superline $\tilde{S}$ describes two--particle fluctuations that have both pure bosonic and collective fermionic character. Filled lines denote the dual propagators, dashed lines the impurity ones.}
\label{fig:Superline}
\end{figure}

Figure~\ref{fig:Superline} 
is a diagrammatic representation of the ``superline'' in Eq.~\eqref{eq:superline}, where the two--particle ladder was inserted for $\tilde{\Pi}$ since we only want to consider two--particle processes. The ladder is necessary for charge conservation~\cite{Hafermann14-2} (see Sec.~\ref{sec:charge}).
These diagrams contain only a dual bosonic line and two--fermionic ladders, so they describe all two--particle processes and no processes involving more than two particles.

\begin{figure}[ht]
\includegraphics[width=0.65\linewidth]{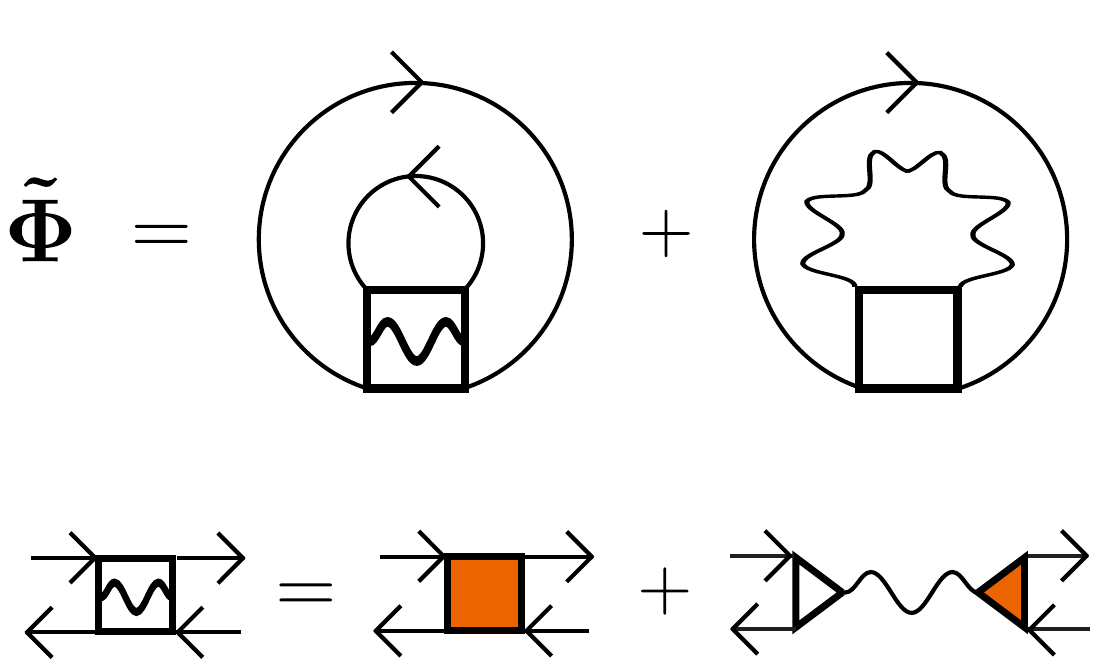}
\caption{(Color online) Dual Luttinger--Ward functional for the Dual Boson approach. The generation of diagrams for the dual self--energy and polarization function occurs by cutting one fermionic or bosonic line respectively.}
\label{fig:Functional}
\end{figure}
\begin{figure}[ht]
\includegraphics[width=0.8\linewidth]{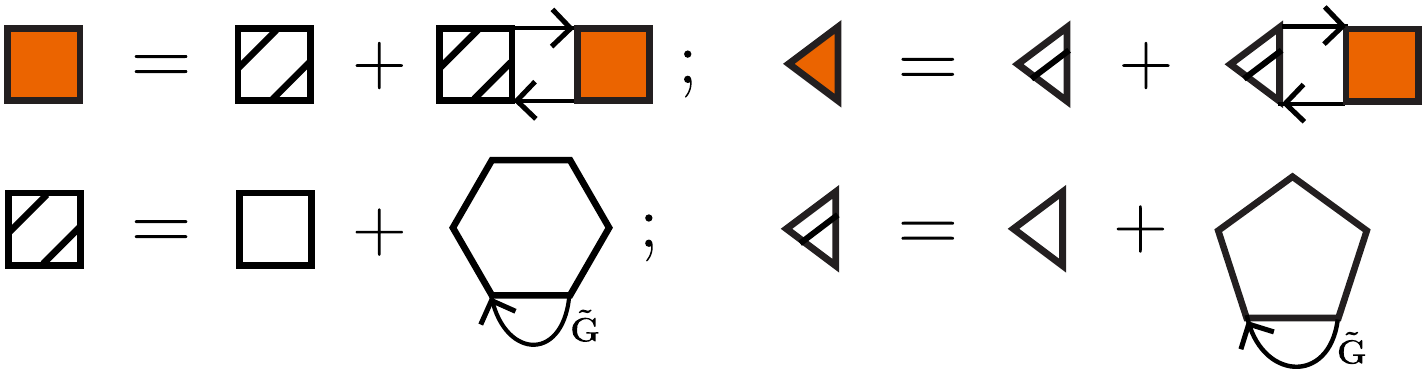}
\caption{(Color online) The renormalized three-- and four--point vertex functions in the Dual Boson approach. The five-- and six--point vertices are introduced here with the local dual fermion line. Such diagrams are identicaly equal to zero because of the self--consistency condition and inserted only to generate the self--energy diagrams from the Luttinger--Ward functional.}
\label{fig:DLines}
\end{figure}

As usual, the diagrams for the fermionic self energy and polarization operator can be found diagrammatically by cutting the respective line of the associated dual functional (see Fig.~\ref{fig:Functional}),
\begin{align}
\tilde{\Sigma} = \frac{\delta\tilde{\Phi}}{\delta\tilde{G}};\,\,\,\,
\tilde{\Pi} = \frac{\delta\tilde{\Phi}}{\delta\tilde{X}}.
\end{align}
There is however an important peculiarity. 
Since the local part of the fermionic propagator is zero due to the self--consistency condition \eqref{eq:sc:df},
there is no need to account for closed fermionic loops in diagrams for the self--energy and polarization operator. 
However, this is not the case for the dual functional diagrams. 
Indeed, cutting a closed loop in the dual functional
yields a non--zero contribution into the local part of the self--energy. 
To account for this, we introduce the renormalized vertices of the dual functional. 
Each renormalized vertex is the sum of a bare part and a higher--order vertex with a closed fermionic loop (see Fig.~\ref{fig:DLines}). 
These renormalized vertices are used as building blocks for the ladders in the functional of Fig.~\ref{fig:Functional}.
This is the minimal possible functional that describes the full physics of two--particle fluctuations.

\begin{figure}[]
\includegraphics[width=0.8\linewidth]{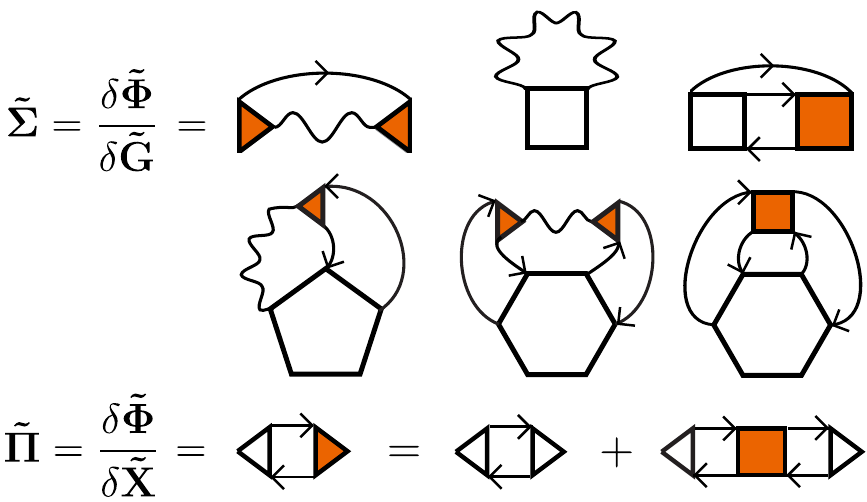}
\caption{(Color online) Diagrams for the dual self--energy and polarization function. The renormalized three-- and four--point vertices have the ladder structure (see Fig.~\ref{fig:DLines}) and don't contain any diagrams with the local fermionic line due to the self--consistency condition.}
\label{fig:SelfEnergy}
\end{figure}

The diagrams for the self--energy and polarization can be determined from the functional. These are shown in Fig.~\ref{fig:SelfEnergy}, with renormalized three-- and four--point ladder vertex functions as in Fig.~\ref{fig:DLines}.

\section{self--consistency condition and higher--order vertices}
\label{HOV}

The dual action contains vertex functions of higher order than the two--particle correlators $\gamma^{2,1}$ and $\gamma^{4,0}$ (see Fig.~\ref{fig:vertices}).
There is no small parameter in the dual perturbation theory, so in principle the contribution of all diagrams containing these additional vertices should be taken into account.
However, this is usually not feasible in practice.
Neglecting higher--order vertices can lead to small but noticeable deviations~\cite{Hafermann09,vanLoon15}.
Although they are usually small, it would be good to reduce these deviations further.

The fermionic self--consistency condition already removes all diagrams with higher order vertices that contain local dual Green's functions. 
In the dual fermion approach, the hybridization function $\Delta_{\nu}$ is the only free parameter, so there is only one self--consistency 
condition, and there is no more freedom to simplify the diagrammatic expansion.
However, the DB approach has an additional free parameter, the retarded interaction $\Lambda_{\omega}$, and it can be chosen self--consistently 
in such way as to minimize the impact of higher--order vertices. 

\begin{figure}[]
\includegraphics[width=0.9\linewidth]{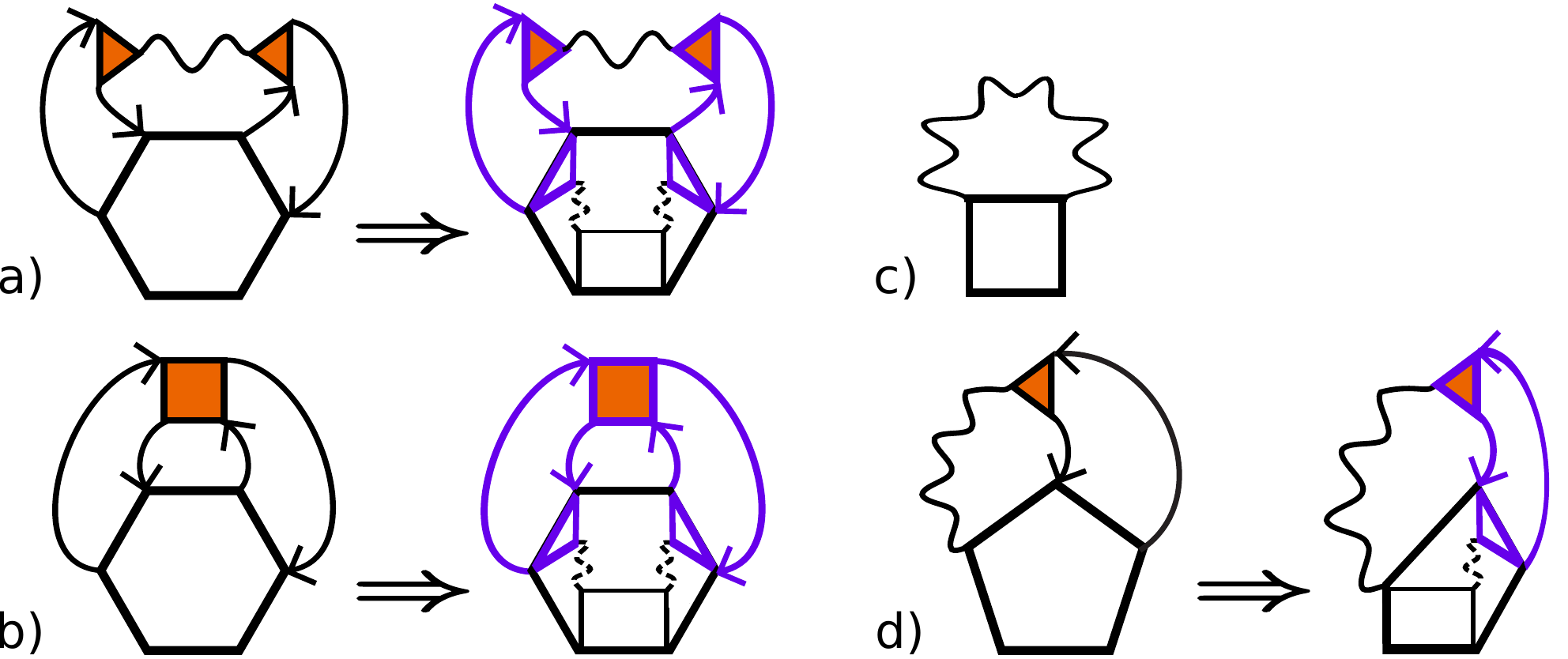}
\caption{(Color online) The structure of the irreducible parts of higher--order vertices with respect to the bosonic arguments. The diagrams a) and b) give the contribution $\tilde{\Sigma}^{6,0}$ to the self--energy from the six--point vertex, diagrams c) and d) are the contributions $\tilde{\Sigma}^{2,2}$ and $\tilde{\Sigma}^{4,1}$ from the vertices $\gamma^{2,2}$ and $\gamma^{4,1}$ respectively. The purple parts of the diagrams are equal to the dual polarization function $\tilde{\Pi}$ (see Fig.~\ref{fig:SelfEnergy}), the dashed wave lines are the impurity susceptibility $\chi$.
}
\label{Transformation}
\end{figure}

We recall that DMFT captures the single--electron hybridization physics and is formulated in terms of the single--particle Green's function. 
The DMFT choice of $\Delta_{\nu}$ removes the lowest--order Hartree
diagram~\cite{Rubtsov09} containing the two--particle quantity $\gamma^{4,0}$. 
Here we would like to include the bosonic (or, equally, two--fermionic) effects. 
In analogy to the foregoing, we construct a theory in terms of single--
and two--particle quantities, and require that the lowest--order diagrams with the three--particle quantities $\gamma^{6,0}$, $\gamma^{4,1}$ and $\gamma^{2,2}$ drop out of the self--energy.
These lowest--order diagrams are shown in Fig.~\ref{Transformation}, where $\tilde{\Sigma}^{n,m}$ denotes the contribution to the dual self energy from the $\gamma^{n,m}$ vertex function.

These contributions to the dual self--energy from higher--order vertices can be divided into reducible and irreducible parts~\cite{Katanin13}.
Here we consider the irreducible contribution from the higher--order vertices; Appendix~\ref{app:12reduc} describes the reducible contributions.

The three--particle vertex $\gamma^{6,0}$ in these diagrams is connected to four internal fermion lines. Counting the number of adjustable parameters, it is easy to realize that a four--time (three frequency) bosonic hybridization,
$\Lambda_{122'1'} c^\dag_1 c^\dag_2 c_{2'} c_{1'}$
is needed to completely remove the diagrams of Fig.~\ref{Transformation}.

Such a quantity is very hard to implement in any practical
calculation. Therefore, we stay with the two--time (one frequency) hybridization $\Lambda_{\omega}$. 
In this way, we cannot make the contribution from the diagrams with three--particle correlators exactly zero. 
However, we will show that the self--consistency condition \eqref{eq:superline} corresponds to removing the contribution from the physically important part of $\gamma^{6,0}$. 
In the case of small $U$, all three--particle correlation effects vanish because of the perturbative form of the diagrammatic expansion of the dual functional (see Appendix~\ref{app:small_red}).

The three--particle vertices in the dual perturbation theory are impurity correlation functions. 
Since the impurity only contains two--particle interaction [see Eq.~\eqref{eq:s:imp}], $\gamma^{6,0}$ can be written as a set of diagrams containing the irreducible two--particle vertex function of the impurity problem connected via local fermionic and bosonic Greens functions. 
Such a diagrammatic expansion of the dual--fermion six--leg vertex was discussed in Ref.~\onlinecite{Katanin13}.

Here we analyze three--particle dual self--energy diagrams where the incoming particle is connected to a two--particle ladder, which depends on a single bosonic frequency.
Fig.~\ref{Transformation} shows these processes. 
Physically, these diagrams describe the interaction of a fermion with a collective excitation. The collective excitation can take the shape of a dual boson propagator or of a ladder of dual fermion propagators.

Let us start with diagrams a) and b). These diagrams give the following contribution to the self--energy, where we denote the purple parts of the diagrams in Fig.~\ref{Transformation} as the dual polarization function $\tilde{\Pi}$: 
\begin{align}
\tilde{\Sigma}^{a}+\tilde{\Sigma}^{b} = \frac12\gamma^{2,2}\chi\big[\tilde{\Pi}+
\tilde{\Pi}\tilde{X}\tilde{\Pi}\big]\chi.
\end{align}
Here $\frac12$ appears as a symmetry factor due to indistinguishable vertices.

The contributions of diagrams c) and d) can be written in the same way,
\begin{align}
\tilde{\Sigma}^{c} &= \frac12\gamma^{2,2}\tilde{X},
\notag \\
\tilde{\Sigma}^{d} &= \gamma^{2,2}\tilde{X}\tilde{\Pi}\chi,
\end{align}
where $\frac12$ in $\tilde{\Sigma}^{c}$ is also a symmetry coefficient.

Then, the total contribution to the dual self--energy from these diagrams is
 \begin{align}
\tilde{\Sigma}^{6} &= \notag \tilde{\Sigma}^{a} +\tilde{\Sigma}^{b}+\tilde{\Sigma}^{c}+\tilde{\Sigma}^{d}
\notag \\
&= \frac12\gamma^{2,2}\big(\tilde{X} + 2\tilde{X}\tilde{\Pi}\chi + \chi\big[\tilde{\Pi}+
\tilde{\Pi}\tilde{X}\tilde{\Pi}\big]\chi \big).
\label{sl0}
\end{align}
Here, we recognize the superline of Eq. \eqref{eq:superline}. This allows to rewrite \eqref{sl0} as
\begin{align}
\tilde{\Sigma}^{6} =\frac12\gamma^{2,2} \tilde{S}_{\text{loc}},
\end{align}
where the local part appears since both ends of the superline are attached to the same vertex.
This expression is equal to zero if we use the self--consistency condition \eqref{eq:superline}.
Therefore, the local part of the ``superline'' $\tilde{S}$ being zero
means that the dual self--energy contribution associated with $\gamma^{2,2}$ drops out of the perturbation theory.

For comparison, the self--consistency condition \eqref{eq:sc:db} corresponds to $\tilde{\Sigma}^{c}=0$. In that case, diagrams a), b) and d) still contribute.

Having removed the lowest--order three--particle processes with three--particle vertices, we can now restrict ourselves to the the simple ladder expansion for the dual self--energy and polarization function
\begin{align}
\tilde{\Sigma}=&
\includegraphics[width=0.2\linewidth]{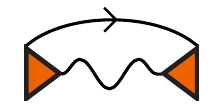}
+
\includegraphics[width=0.2\linewidth]{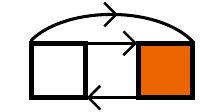} 
\label{eq:sigma}
\\
\tilde{\Pi}=&
\includegraphics[width=0.14\linewidth]{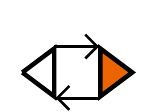}.
\label{eq:pi}
\end{align}
The reducible contributions to the three--particle vertexes can be accounted 
by the re--definition on the lines in these diagrams, as it is described in Appendix~\ref{app:12reduc}. These expressions allow to close the set of equations for dual Green's functions formed by 
the self--consistency conditions  (\ref{eq:sc:df}, \ref{eq:superline}) and definitions (\ref{eq:barefermionpropagator}, \ref{eq:barebosonpropagator}).

Thus, we obtained a diagrammatic argument for the bosonic self--consistency similar to that in the DF approach. 
This condition removes the physically important dual self--energy contribution from higher--order vertices and simplifies the diagrammatic expansion of the interaction functional $\tilde{V}$. 
The main difference is that for the bosonic sector, both single--boson and two--fermion contributions are crucial. 
 
It is important to note that a nonlocal self--energy can break the
charge conservation law, since the vertex obtained as a second
derivative of the dual Luttinger--Ward functional then depends on three
momenta, whereas the ladder approximation only takes one transferred
momentum into account.
For this reason, in this work we restrict ourselves to a local self--energy and focus on nonlocal bosonic effects of Eq.~\eqref{eq:pi}. 
As a result, the two self--consistency conditions \eqref{eq:sc:dmft} and \eqref{eq:sc:df} for the fermionic hybridization are equivalent.
The fermionic and bosonic self--consistency conditions remove the main physical contributions to $\tilde{\Sigma}$ and the fermion physics is accounted for on the level of the impurity.

\begin{figure}[]
 \includegraphics{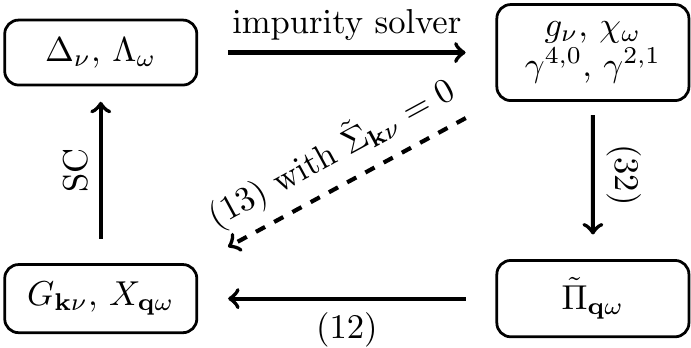}
 \caption{A summary of the computational scheme used. The impurity solver determines local impurity quantites based on hybridization functions $\Delta_\nu,\Lambda_\omega$. The fermionic lattice Green's function is determined directly from the impurity quantities. For the lattice susceptibility, we take additional nonlocal corrections into account via the dual polarization $\tilde{\Pi}$ in the ladder approximation.}
 \label{fig:flowchart}
\end{figure}

Our computational scheme is summarized in Figure~\ref{fig:flowchart}. 
We start with an initial guess for the hybridization functions $\Delta_\nu$ and $\Lambda_\omega$. 
In this work we take the result of a converged EDMFT simulation as the inital guess, however it is also possible to start from non--interacting Green's functions or from a previously obtained converged DB result at other parameters. 
The impurity problem is solved using quantum Monte Carlo~\cite{Rubtsov05,Werner06,Hafermann13,Hafermann14} and the resulting impurity quantities are used to calculate the dual polarization. Then, the lattice Green's function and susceptibilty are determined and new hybridization functions are determined using the self--consistency conditions. This procedure is iterated until convergence is achieved.

\section{Charge conservation}
\label{sec:charge}

Here, we prove that the susceptibility obtained from DB satisfies the charge conservation requirement as long as the bare (local) self--energy and the local vertices are used. The conditions for the proof are:
\begin{itemize}
 \item The impurity problem \eqref{eq:s:imp} is solved exactly, so that the impurity vertex and self--energies satisfy the Ward identity.
 \item The self--energy of the fermions is local and equal to the impurity self--energy, $\Sigma_{\kv\nu}=\Sigma^{\text{imp}}_{\nu}$. 
 \item The hybridization $\Delta_\nu$ is chosen in such a way that the local part of the dual fermion Green's function is zero, i.e., the self--consistency condition \eqref{eq:sc:df} is satisfied.
\end{itemize}
It is important to note that for this proof there is no condition on the retarded interaction $\Lambda_\omega$. In particular, the proof works for the self--consistent choice of $\Lambda_\omega$ in the previous section.

We start with the dual polarization in the ladder approach~\cite{vanLoon14-2}, Eq.~\eqref{eq:pi}, and write it as
\begin{align}
 \tilde{\Pi}_{\qv\omega} =& \sum_{\nu} \gamma^{2,1}_{\nu+\omega} \tilde{B}_{\qv\nu\omega} \Gamma^{2,1}_{\qv\nu\omega},
\end{align}
where we have defined the bubble of dual Green's functions
\begin{align}
 \tilde{B}_{\qv\nu\omega} =& -\sum_{\kv} \tilde{G}_{\kv\nu}\tilde{G}_{\kv+\qv\nu+\omega},
\end{align}
and renormalized fermion--boson and fermion--fermion vertices
\begin{align}
\Gamma^{2,1}_{\qv\nu\omega} &= \gamma^{2,1}_{\nu\omega} - \sum_{\nu'} \Gamma^{4,0}_{\qv\nu\nu'\omega} \tilde{B}_{\qv\nu'\omega} \gamma^{2,1}_{\nu'\omega}
\label{eq:G21r}
\\
\Gamma^{4,0}_{\qv\nu\nu'\omega} &= \gamma^{4,0}_{\nu\nu'\omega} - \sum_{\nu''} \gamma^{4,0}_{\nu\nu''\omega} \tilde{B}_{\qv\nu\omega} \Gamma^{4,0}_{\qv\nu''\nu'\omega}.
\end{align}
The results of the ladder summation can be expressed as a geometric series,
\begin{align}
 \Gamma^{2,1}_{\qv\nu\omega} =& \sum_{\nu''} \left[ I+\tilde{B}_{\qv\nu''\omega} \gamma^{4,0}_{\nu\nu''\omega} \right]^{-1} \gamma^{2,1}_{\nu''\omega},\notag \\
 \Gamma^{4,0}_{\qv\nu\nu'\omega} =& \sum_{\nu''} \left[ I+\tilde{B}_{\qv\nu''\omega} \gamma^{4,0}_{\nu\nu''\omega} \right]^{-1} \gamma^{4,0}_{\nu''\nu\omega}.
\end{align}
where the inverse should be understood as a matrix inversion in frequency space and $I$ is the identity matrix in this space $(\nu\nu'')$.

Now, we proceed by going from the impurity vertex to the particle--hole irreducible impurity vertex (see Refs.~\onlinecite{Toschi07,Katanin09}).
\begin{align}
B^{\text{imp}}_{\nu\omega} =& - g_\nu g_{\nu+\omega}, \notag \\
\gamma^{4,0}_{\nu\nu'\omega} =& \sum_{\nu''} \left[I + \Gamma^{4,0\,\text{ir}}_{\nu\nu''\omega} B^{\text{imp}}_{\nu\omega}\right]^{-1} \Gamma^{4,0\,\text{ir}}_{\nu''\nu'\omega}, \label{eq:gammair}\\
\gamma^{2,1}_{\nu \omega } =&\chi _{\omega }^{-1} \sum_{\nu''}
\left[ I+\Gamma _{\nu \nu ^{\prime \prime }\omega }^{4,0\,\mathrm{ir}} B _{\nu
^{\prime \prime }\omega }^{\text{imp}}\right] ^{-1}.
\end{align}

The impurity susceptibility $\chi_\omega$ is related to this irreducible vertex as
\begin{align}
 \chi_\omega =& \sum_{\nu\nu''} \left[ I+\Gamma^{4,0\,\text{ir}}_{\nu\nu''\omega}
 B^{\text{imp}}_{\nu''\omega} \right]^{-1} B^{\text{imp}}_{\nu\omega}.\label{eq:chiir}
\end{align}

Now, all relevant two--particle correlators of the impurity have been expressed in terms of two quantities, the irreducible vertex $\Gamma^{4,0\,\text{ir}}$ and the particle--hole bubble $B^{\text{imp}}$. This allows us to write the combination $P=-\chi_\omega-\chi_\omega\tilde{\Pi}_{\qv\omega}\chi_\omega$ that enters the lattice susceptibility in Eq. \eqref{eq:xfrompid} entirely in terms of these two quantities.
\begin{align*}
 P_{\qv\omega} =& -\chi_\omega-\chi_\omega\tilde{\Pi}_{\qv\omega}\chi_\omega \notag\\
 =& - \chi_\omega - \sum_{\nu_1\nu_2\nu_3\nu_4} \left[I+B^{\text{imp}}_{\nu_2\omega} \Gamma^{4,0\,\text{ir}}_{\nu_2\nu_1\omega} \right]^{-1} \tilde{B}_{\qv\nu_1\omega} \notag \\
 &\times \left[I+\gamma^{4,0}_{\nu_1\nu_3\omega} \tilde{B}_{\qv\nu_3\omega}\right]^{-1}
 \left[ I + \Gamma^{4,0\,\text{ir}}_{\nu_3\nu_4\omega} B^{\text{imp}}_{\nu_4\omega}\right]^{-1} \notag \\
 \overset{\eqref{eq:gammair}}{=}&- \chi_\omega - \sum_{\nu_1\nu_2\nu_4} \left[I+B^{\text{imp}}_{\nu_2\omega} \Gamma^{4,0\,\text{ir}}_{\nu_2\nu_1\omega} \right]^{-1} \tilde{B}_{\qv\nu_1\omega} \notag \\
 &\times
 \left[ I + \Gamma^{4,0\,\text{ir}}_{\nu_1\nu_4\omega} (B^{\text{imp}}_{\nu_4\omega}+\tilde{B}_{\qv\nu_4\omega})\right]^{-1} \notag \\
 \overset{\eqref{eq:chiir}}{=}&- \sum_{\nu_1\nu_2\nu_4} \left[I+B^{\text{imp}}_{\nu_2\omega} \Gamma^{4,0\,\text{ir}}_{\nu_2\nu_1\omega} \right]^{-1} \notag \\
 &\times
 \left[ B^{\text{imp}}_{\nu_1\omega} \delta_{\nu_1\nu_4} + \tilde{B}_{\qv\nu_1\omega}
 \left[ I + \Gamma^{4,0\,\text{ir}}_{\nu_1\nu_4\omega} B_{\qv\nu_4\omega}\right]^{-1} \right] \notag \\
 \overset{\eqref{eq:chiir}}{=}&- \sum_{\nu_1\nu_2\nu_3\nu_4} \left[I+B^{\text{imp}}_{\nu_2\omega} \Gamma^{4,0\,\text{ir}}_{\nu_2\nu_1\omega} \right]^{-1} \notag \\
 &\times
 \left[ B^{\text{imp}}_{\nu_1\omega} \left[ I + \Gamma^{4,0\,\text{ir}}_{\nu_1\nu_3\omega} B_{\qv\nu_3\omega}\right] + \tilde{B}_{\qv\nu_1\omega} \delta_{\nu_1\nu_3}
 \right]  \notag \\
  &\times
  \left[ I + \Gamma^{4,0\,\text{ir}}_{\nu_3\nu_4\omega}
 B_{\qv\nu_4\omega}\right]^{-1} \notag \\
\end{align*}
\begin{align}
 =&- \sum_{\nu_1\nu_2\nu_3\nu_4} \left[I+B^{\text{imp}}_{\nu_2\omega} \Gamma^{4,0\,\text{ir}}_{\nu_2\nu_1\omega} \right]^{-1} \notag \\
 &\times
 \left[ I + B^{\text{imp}}_{\nu\omega} \Gamma^{4,0\,\text{ir}}_{\nu_1\nu_3\omega} \right] \times B_{\qv\nu_3\omega}
   \notag \\
  &\times
  \left[ I + \Gamma^{4,0\,\text{ir}}_{\nu_3\nu_4\omega}
 B_{\qv\nu_4\omega}\right]^{-1} \notag \\
 =&- \sum_{\nu_1\nu_2} B_{\qv\nu_1\omega}
  \left[ I + \Gamma^{4,0\,\text{ir}}_{\nu_1\nu_2\omega}
 B_{\qv\nu_2\omega}\right]^{-1}.
\end{align}
In the Hubbard model ($V=0$), this is the expression for the DMFT susceptibility.
Here we have introduced the lattice bubble
\begin{align}
 B_{\qv\nu\omega} =&  B^{\text{imp}}_{\nu\omega} +  \tilde{B}_{\qv\nu\omega} \notag \\
 =&\sum_{\kv} G_{\kv\nu}G_{\kv+\qv,\nu+\omega}.
\end{align}
This relation holds, since $G_{\kv\nu}=g_\nu+\tilde{G}_{\kv\nu}$ and the local part of the dual Green's function is zero so cross--terms $g\tilde{G}$ vanish in the momentum sum.

For the charge vertex
\begin{align}
\Gamma _{\mathbf{q}\nu \omega }^{4,0\,\mathrm{ch}}=\sum_{\nu'}\left[
I+\Gamma _{\nu \nu ^{\prime }\omega }^{4,0\,\mathrm{ir}}B_{\mathbf{q,}\nu
^{\prime }\omega }\right] _{\nu \nu ^{\prime }}^{-1}
\end{align}%
we have the Ward identity~\cite{Hafermann14-2}
\begin{align}
\omega \Gamma _{\qv=0\nu \omega }^{4,0\,\mathrm{ch}}=G_{\mathbf{k},\nu
}^{-1}-G_{\mathbf{k},\nu +\omega }^{-1}=-(\omega -\Sigma _{\nu +\omega
}+\Sigma _{\nu }).
\end{align}%
Therefore, for $\omega\neq 0$,%
\begin{align}
P _{\qv=0\omega }=&
-\sum_{\nu} B_{\mathbf{q,}\nu \omega
}\Gamma _{\qv=0\nu \omega }^{4,0\,\mathrm{ch}} 
\notag \\
&= - \sum_{\nu} \sum_{\kv} G_{\kv\nu}G_{\kv,\nu+\omega} [G_{\mathbf{k},\nu
}^{-1}-G_{\mathbf{k},\nu +\omega }^{-1}] \notag \\
&= \sum_\nu \sum_\kv [G_{\kv\nu}-G_{\kv,\nu+\omega}] \notag \\
&= 0.
\end{align}%
The physical polarization is given by%
\begin{align}
\Pi _{\mathbf{q}\omega }^{-1}=P _{\mathbf{q}\omega }^{-1}-\Lambda
_{\omega },
\end{align}%
which yields the required condition $\Pi _{\qv=0,\omega>0}=0$. For arbitrary
$\mathbf{q}$ the physical susceptibility is given by%
\begin{align}
X_{\mathbf{q}\omega }=\frac{1}{-\Pi _{\mathbf{q}\omega }^{-1}-V_{q}-U}=-%
\frac{1}{P _{\mathbf{q}\omega }^{-1}+U+V_{\mathbf{q}}-\Lambda _{\omega
}},
\end{align}%
which gives
\begin{align}
 X_{\qv=0,\omega\neq0} = 0.
\end{align}
So the charge conservation requirement is satisfied.

\section{Application of the self--consistency condition}
\label{sec:application}

Having introduced a new self--consistency condition \eqref{eq:sc:edmft}, we will study its behavior in some important limits.
In the limit of infinite dimensions, DB with the new self--consistency condition reduces to the (exact in $d=\infty$) DMFT.
In the non--interacting limit, the new self--consistency condition is automatically satisfied by the exact solution.
At weak interaction, the self--consistency condition changes the effective impurity interaction.
Finally, we determine the high--frequency limit of the retarded interaction and find that it is in general nonzero.

\subsection{Limit of infinite dimensions}

As is well known, DMFT becomes exact in the limit of infinite dimensions~\cite{Metzner89,Georges96}. Our DB scheme inherits this property. 
To prove this, it is only necessary to show that DB and DMFT are equivalent, i.e., that the self--consistency condition \eqref{eq:sc:edmft} is 
satisfied for vanishing retarded interaction in the limit of infinite dimensions. Since nonlocal interactions only contribute at the Hartree 
level in infinite dimensions~\cite{Georges96}, it is sufficient to prove this for $V_\qv=0$.
\begin{align}
 X_{\qv\omega} \overset{\Lambda_\omega=0}= &\chi_{\omega} + \chi_{\omega}
 \tilde{\Pi}_{\qv\omega} \chi_\omega \notag \\
 \sum_\qv  X_{\qv\omega} - \chi_\omega = \,\,\,\,&\chi_\omega \sum_{\qv} \tilde{\Pi}_{\qv\omega} \chi_\omega \notag \\ 
 = \,\,\,\,&\chi_\omega \sum_{\qv} \tilde{\Pi}^{(2)}_{\qv\omega} \chi_\omega + \chi_\omega \sum_{\qv} \tilde{\Pi}^{(>2)}_{\qv\omega} \chi_\omega \notag \\
 = \,\,\,\,&0, 
\label{eq:infinited}
\end{align}
since the local part of the second--order diagram vanishes due to the fermionic self--consistency (as before), and the local part of higher--order 
vertex corrections vanishes in infinite dimensions~\cite{Khurana90}.
The vanishing of the higher--order diagram can also be understood in a $1/d$ expansion~\cite{Metzner89,Georges96}: the four dual Green's functions 
contribute $\mathcal{O}(1/\sqrt{d})$ each~\cite{Otsuki14}, and there is a single internal summation contributing $\mathcal{O}(d)$, so 
$\tilde{\Pi}^{(>2)}_{\qv\omega} \propto \mathcal{O}(1/d)$.

Eq. \eqref{eq:infinited} shows that the self--consistency condition \eqref{eq:sc:edmft} is automatically satisfied for $\Lambda_\omega = 0$ in infinite 
dimensions, so DB and DMFT are equivalent (and both exact) in this limit.

\subsection{Noninteracting system}
\label{sec:noninteracting}

The noninteracting system ($U=0$, $V=0$) is exactly solvable, the bare Green's function is the exact lattice Green's function. This result is recovered in self--consistent DB if $\Lambda_{\omega}=0$. Here, we will show that the new self--consistency condition is indeed fulfilled when $U=0$ and $\Lambda_\omega =0$. Starting with
a noninteracting impurity problem gives $\gamma^{(4)}_{\nu\nu'\omega} = 0$, and consequently~\cite{Rubtsov12} $\lambda_{\nu\omega} = -1/\chi_\omega$. Only the second--order
diagram \begin{align}
 \chi_\omega\tilde{\Pi}^{(2)}_{\qv,\omega}\chi_\omega &= [\tilde{G}\tilde{G}]_{\qv,\omega}
\end{align}
contributes to the dual polarization, with $\tilde{G}$ the Green's function of the dual fermions. Its local part vanishes due to the self--consistency condition
on the fermions, and as a result the local part of the bubble $[\tilde{G}\tilde{G}]_{\qv,\omega}$ is also zero. So the local part of the 
lattice susceptibility \eqref{eq:xfrompid} is
\begin{align}
 \sum_{\qv} X_{\qv,\omega} &= \sum_{\qv} \chi_{\omega} + \sum_\qv [\tilde{G}\tilde{G}]_{\qv,\omega} \\
 &= \chi_{\omega},
\end{align}
and the self--consistency condition \eqref{eq:sc:edmft} is satisfied.

\subsection{Strong and weak interaction}
\label{sec:strongweak}

In the special case of the Hubbard model (that is, $V=0$), the interaction is purely local, but the collective modes are still present. Magnons are known to determine the low--energy physics of the half--filled Hubbard model.

The dual fermion ladder summation allows us to describe the effect of magnon fluctuations, leading to the formation of an ``antiferromagnetic'' pseudogap slightly above the Neel temperature~\cite{Hafermann09}. 
In the same vein, one should expect the presence of collective density modes. 
Indeed, it is widely accepted that the sligtly doped Hubbard model shows instabilities related to the formation of static density waves. 
This indicates the importance of dynamical density fluctuations in a wider region, possibly including the case of half--filling.

In the dual--boson theory, the presence of collective modes should be reflected in a corresponding bosonic hybridization $\Lambda$.
Indeed, our self--consistency conditions impose that $\Lambda$ is non--zero even for $V=0$. In this case the bare bosonic line is purely local, and the nonlocality is provided by the fermionic ladder. The latter exactly corresponds to a collective mode, as it was discussed above.

Two important limits of the Hubbard model ($V=0$) are those of small and large $U$. Both of these limits are described by $\chi_\omega \Lambda_\omega \ll 1$:
for $U = 0$ the $\Lambda_\omega = 0$ as it was shown in Sec.~\ref{sec:noninteracting}, for $U\rightarrow 0$ the $\Lambda_\omega \rightarrow 0$ while $\chi_\omega$ stays finite.
Conversely, for large $U$, the susceptibility $\chi_\omega$ becomes very small.
This allows us to expand the charge susceptibility \eqref{eq:xfrompid} in $\chi_\omega\Lambda_\omega$
\begin{align}
 X_{\qv\omega}
 \approx&
 \chi_{\omega} + \chi_{\omega}\tilde{\Pi}_{\qv\omega}\chi_{\omega} - \chi_\omega \Lambda_\omega\chi_\omega + \ldots \label{eq:strongweak1}
\end{align}
The polarization also simplifies in the limits of both strong and weak interaction. With the same argument as in DF~\cite{Rubtsov08}, the vertex $\gamma$ is 
small at small $U$, whereas the dual Green's function $\tilde{G}$ is small at large $U$. This means that both at large and at small $U$, higher orders 
in the ladder decay quickly, and when combining the self--consistency condition \eqref{eq:sc:edmft} and Eq.~\eqref{eq:strongweak1}, only the third--order 
contribution $\tilde{\Pi}^{(3)}$ needs to be considered (the local part of the second--order diagram drops out due to the self--consistency condition on the fermions, as before).
\begin{align}
  \chi_\omega \Lambda_\omega\chi_\omega \label{eq:strongweak2}
  \approx&
  \sum_{\qv} \chi_{\omega}\tilde{\Pi}^{(3)}_{\qv\omega}\chi_{\omega}.
\end{align}

\subsection{Symmetries in the Hubbard model}
\label{sec:symmetries}

At small $U$, the third--order polarization diagram is proportional to $U$, since it contains a single vertex $\gamma$.
In this work, we focus on the charge sector, however, there are also collective excitations in the magnetic sector and the DB formalism can 
be also applied to $S_z S_z$ interactions. 
An important difference between the charge and magnetic channel is the sign of the effective interaction.
This implies, to lowest order in $U$, opposing signs in the third--order dual polarization 
$\tilde{\Pi}^{(3),\text{charge}}_{\qv\omega}=-\tilde{\Pi}^{(3),\text{magnetic}}_{\qv\omega}$ and effective impurity interaction
$\Lambda_\omega^{\text{charge}}+\Lambda_\omega^{\text{magnetic}}=0$.

By using particle-hole symmetry, one can show (see, e.g., chapter 11 of Ref.~\onlinecite{julich2015} for a pedagogical discussion) that the transformation $U\rightarrow -U$ in the half-filled square lattice Hubbard model interchanges charge and spin fluctuations. 
Single-particle properties like the self-energy are invariant under the the transformation $U\rightarrow -U$.
Introducing a self-consistency only on the charge sector, as we do here, will break this symmetry. To retain the $U\rightarrow -U$ symmetry, the charge and magnetic channel need to be treated on the same footing. 
To also retain spin rotation symmetry, and in fact the full $SO(4)$ symmetry~\cite{julich2015}, all three magnetic channels are needed, as well as the charge channel and two superconducting channels. 

\subsection{Screening by nonlocal interaction}

Nonlocal charge fluctuations play an especially important role in systems with nonlocal interactions. 
The nonlocal interaction can cause quantitative differences by screening the effective local interaction.
However, it can also lead to qualitatively different physics.
The checkerboard order that arises from repulsive nearest-neighbor interaction is a good example of this, that has been studied both in EDMFT~\cite{Ayral13,Huang14} and in dual boson~\cite{vanLoon14-2}.
Attractive nearest-neighbor interaction, on the other hand, can lead to phase separation into high and low density phases~\cite{Gubernatis85}. 
Clearly, the sign of  the nearest-neighbor interaction $V$ is physically relevant.
Using the Bogoliubov inequality, screening effects have been estimated to be proportional to $V$~\cite{Schuler13}.

A peculiarity occurs for EDMFT on the square lattice. As shown in Appendix~\ref{app:edmft:fail}, due to the structure of the momentum sums, only the absolute value of $V$ matters to EDMFT.
In dual boson, on the other hand, the sign of $V$ does matter.

\subsection{Numerical results at weak interaction}

\begin{figure}
 \includegraphics[]{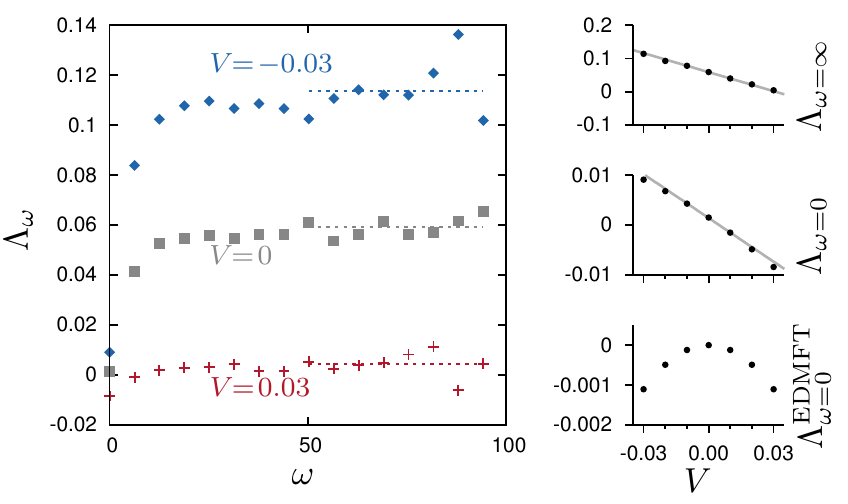}
 \caption{(Color online) Left: Effective impurity interaction as a function of Matsubara frequency for various nonlocal interaction strengths $V$, at $U=0.5$. The dashed line indicates the high--frequency asymptotic value of $\Lambda_\omega$, and this value is shown in the top right as a function of $V$ (the line is a linear fit for $\abs{V}\leq 0.01$.). Center--right is the zero--frequency $\Lambda_\omega$ and bottom right the zero--frequency $\Lambda_\omega$ in EDMFT.
 \label{fig:lowU:multi}
 }
\end{figure}

To start our discussion of the numerical results, we look at a system where correlation effects are expected to be moderate. 
We take a weak local interaction $U=0.5$ and relatively high temperature $T=1$, and vary the nearest--neighbor interaction between $V=-0.03$ and $V=0.03$. 
All parameters are given in units of $t=1$.
The effective impurity interaction after the dual self--consistency is given in Fig.~\ref{fig:lowU:multi}.

One thing that is immediately noticeable from Fig.~\ref{fig:lowU:multi}. is that the high--frequency asymptotic value of the effective impurity interaction depends on the nonlocal interaction strength. This is unlike EDMFT, where the asymptotic is given by the local interaction strength only (see e.g. Fig. 9 of~\onlinecite{Sun02} or Fig. 6c of~\onlinecite{Ayral13}). Fig.~\ref{fig:lowU:multi}b shows that the asymptotic value depends linearly on $V$. This is reminiscent of the effective local interaction $U^\ast=U-\overline{V}$ obtained using the Peierls--Feynman--Bogoliubov variational principle~\cite{Schuler13}, where $\overline{V}$
is proportional to $V$. The linear fit in Fig.~\ref{fig:lowU:multi}b gives $\overline{V}\approx 1.9 V$ in this case.

Alternatively, the zero--frequency part of the retarded interaction can be used to estimate the effective interaction. As mentioned before, in EDMFT this leads to a quadratic dependence~\cite{Huang14} of the effective interaction on $V$. We have observed the same quadratic behavior in EDMFT, see Fig.~\ref{fig:lowU:multi}d. 
The zero--frequency part of the retarded interaction in self--consistent DB is shown in Fig.~\ref{fig:lowU:multi}c. We find that the zero--frequency part of the retarded interaction also depends linearly on $V$, with $\overline{V}\approx 0.3 V$. This constant of proportionality is almost an order of magnitude lower than that of the high--frequency retarded interaction. On the other hand, the zero--frequency retarded interaction in DB (Fig.~\ref{fig:lowU:multi}c) is an order of magnitude larger than in EDMFT (Fig.~\ref{fig:lowU:multi}d).

Fig.~\ref{fig:lowU:multi} shows that even at $V=0$ (Hubbard model), the self--consistency condition introduces a retarded interaction to the impurity problem. In the absence of a nonlocal interaction, the susceptibility still has a nontrivial momentum--dependence and the impurity feels the effect of these nonlocal charge fluctuations.
An important remark here is that, according to Sec. \ref{sec:strongweak}, the retarded interactions in the charge and magnetic channel cancel to first order in $U$ in the Hubbard model. So the finite retarded charge interaction is mostly compensated by the magnetic interaction.

As an aside, the charge conservation requirement $X_{\qv=0,\omega>0}=0$ is satisfied numerically in these self--consistent DB calculations (not shown).

\subsection{Numerical results at strong interaction}

\begin{figure}
 \includegraphics[]{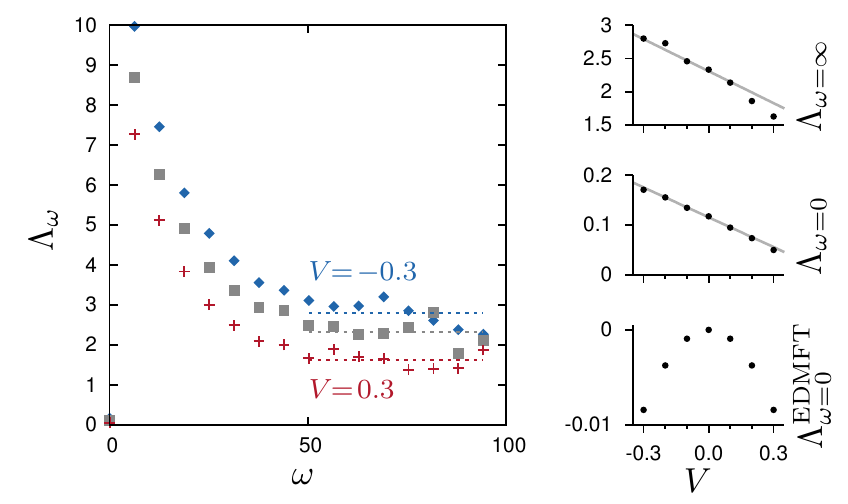}
 \caption{(Color online) Left: Effective impurity interaction as a function of Matsubara frequency for various nonlocal interaction strengths $V$, at $U=10$. The dashed line indicates the high--frequency asymptotic value of $\Lambda_\omega$, and this value is shown in the top right as a function of $V$ (the line is a linear fit for $\abs{V}\leq 0.1$.). Center--right is the zero--frequency $\Lambda_\omega$ and bottom right the zero--frequency $\Lambda_\omega$ in EDMFT.
 \label{fig:highU:multi}
 }
\end{figure}

Now, we turn to a strongly interacting system, with $U=10$, and local interaction varying between $V=-0.3$ and $V=0.3$. The retarded interaction is shown in Fig.~\ref{fig:highU:multi}. As before, there is a constant retarded interaction in the high--frequency limit, the linear fit of this coefficient in Fig.~\ref{fig:highU:multi}b gives $\overline{V}\approx 1.9 V$. The zero--frequency retarded interaction also depends linearly on $V$ with coefficient $\overline{V}\approx 0.2 V$.

Note that the retarded interaction is quite large and positive (repulsive), so the effect of the nonlocal charge fluctuations is to make the system more insulating.

The retarded interaction has a peculiar shape, with a strong repulsion at finite frequency and a much smaller value at zero frequency. The retarded interaction is based on the lattice susceptibility, which has a discontinuity at $\qv=0$ and $\omega=0$.
In that sense, it is not completely surprising that the retarded interaction also changes sharply around zero frequency.
However, one should keep in mind that the results shown here only consider fluctuations in the density channel. As explained in Sec.~\ref{sec:symmetries}, a full computation also needs to take the magnetic and superconducting fluctuations into account and we expect these fluctuations to be very important for strong interaction strengths.

\begin{figure}
 \includegraphics[]{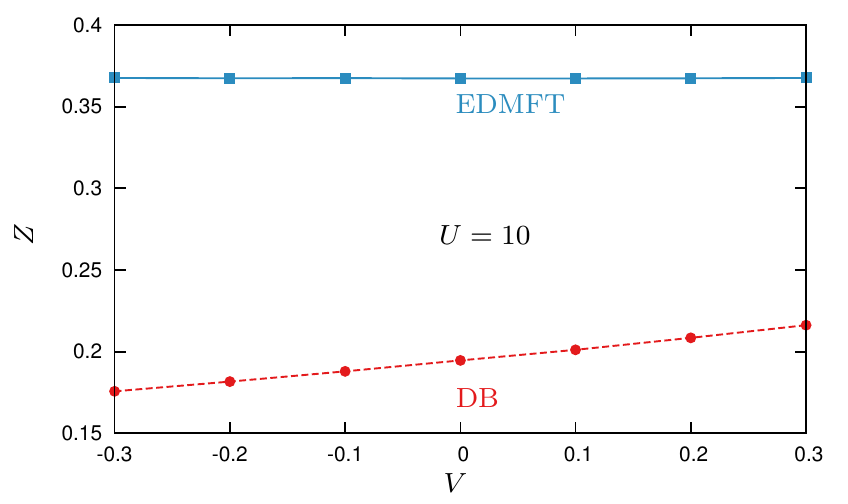}
 \caption{(Color online) Quasiparticle renormalization factor $Z$ as a function of the nonlocal interaction strength $V$, at $U=10$. Nearest-neighbour repulsion ($V>0$) screens the local interaction and increases the quasiparticle renormalization factor.
 \label{fig:highU:Z}
 }
\end{figure}

The change in the impurity interaction affects the local observables obtained from the impurity. 
As an example, in Fig.~\ref{fig:highU:Z}, we show the quasiparticle renormalization factor $Z=(1 - \Imag \Sigma_{\nu_1}/\nu_1)^{-1}$ determined from the local self-energy at the first Matsubara frequency.
The lattice self-consistency condition takes into account feedback of the nonlocal charge correlation effects onto the impurity. 
These correlations drive the system in the direction of the insulating phase and this is reflected in the reduced $Z$-factor.
Secondly, a repulsive nonlocal interaction $V>0$ screens the local impurity problem, makes the system less correlated and increases the quasiparticle weight. 
Within the EDMFT approximation, the dependence on $V$ is a lot weaker.

\begin{figure}
 \includegraphics[]{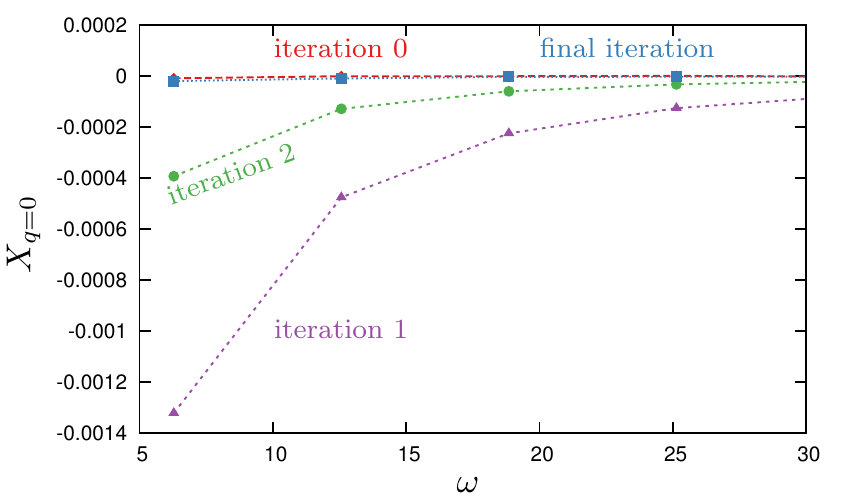}
 \caption{(Color online) Charge conservation requirement $X_{\qv=0,\omega>0}=0$ during the self--consistency scheme. Iteration 0 is single--shot DB. These results are at $U=10$, $V=0.3$.
 \label{fig:highU:charge}
 }
\end{figure}

The charge conservation is also visible numerically. The susceptibility in single--shot DB at the end of EDMFT self--consistency is known to satisfy the charge conservation requirement~\cite{Hafermann14-2} $X_{\qv=0,\omega>0}=0$. This is shown in Fig.~\ref{fig:highU:charge} as iteration 0. Now, when the additional DB self--consistency is started, and $\Lambda_\omega$ is updated, this charge conservation is initially broken, as can be seen in iteration 1 and 2 of Fig.~\ref{fig:highU:charge}. Charge conservation is restored as self--consistency is achieved.

\subsection{High frequency asymptotic of $\Lambda_{\omega}$}

In the previous section, it was shown that the retarded interaction $\Lambda_{\omega}$ goes to a constant at high frequency.
This high frequency asymptotic behavior can be understood in the
following way. 
We start by examining the ingredients of the self--consistency formula. 
The impurity susceptibility $\chi_\omega$ decays as $\omega^{-2}$.
At high frequency, the fermion--boson vertex increases as $\lambda_{\nu\omega}\propto \chi_\omega^{-1}\propto \omega^2$. 
A bubble of dual Green's functions $\tilde{B}_{\qv,\omega}$ decays as $\omega^{-2}$, so the second--order diagram $\tilde{\Pi}^{(2)}_{\qv,\omega} \propto \lambda_\omega \tilde{B}_{\qv,\omega} \lambda_\omega \propto \omega^2$.
This results in a constant $\chi_\omega \tilde{\Pi}^{(2)}_{\qv,\omega}$, which is necessary for the charge conservation condition
$\chi_\omega \tilde{\Pi}^{(2)}_{\qv=0,\omega\rightarrow \infty}=-1$.

So the magnitude of the second--order diagram is increases as a function of frequency. However, here we only consider its local part, which vanishes 
since the local part of the bubble of dual Green's functions is exactly zero due to the fermion self--consistency condition, i.e.,
\begin{align}
 \sum_\qv \tilde{\Pi}^{(2)}_{\qv,\omega} = 0.
\end{align}
The higher order ladder diagrams have additional rungs $\gamma_\omega \tilde{B}_{\omega}$. Since $\gamma$ goes to a constant ($U^{\text{impurity}}$) 
at large frequency, every additional order in the ladder decays by an extra factor $\omega^{-2}$ coming from the bubble.
That means that the first non--vanishing contribution to the local part of $\tilde{\Pi}$ comes from the third order diagram, which will give a constant.

Now we go from the polarization to the lattice susceptibility \eqref{eq:xfrompid} and expand it with respect to $\chi_{\omega}(\Lambda_{\omega}-V_{q}) \ll 1$. 
This is valid at high frequencies, since $\chi_\omega$ decays as $\omega^{-2}$ and we assume $\Lambda_\omega$ does not increase as a function of $\omega$, which will be justified a posteriori:
\begin{align}
X_{q,\omega}=\chi_{\omega}+\chi_{\omega}\tilde{\Pi}_{q,\omega}\chi_{\omega}-(\chi_{\omega}+\chi_{\omega}\tilde{\Pi}_{q,\omega}\chi_{\omega})^2(\Lambda_{\omega}-V_{q}).
\end{align}
Applying the self--consistency condition \eqref{eq:sc:db} gives
\begin{align}
0 =& \sum_{\qv} X_{\qv,\omega} - \chi_\omega \notag \\
0 =& \sum_{\qv} \left[\tilde{\Pi}_{\qv,\omega} - (1+\chi_\omega\tilde{\Pi}_{\qv,\omega})^2 (\Lambda_\omega-V_\qv) \right]\notag \\
0 =& \sum_\qv \tilde{\Pi}^{(2)}_{\qv,\omega} + \sum_\qv \tilde{\Pi}^{(>2)}_{\qv,\omega} \notag \\
& - \sum_\qv (1+\chi_\omega\tilde{\Pi}_{\qv,\omega})^2 (\Lambda_\omega-V_\qv).
\end{align}
According to the previous discussion, the first term vanishes, the second term goes to a constant at large frequency, and $1+\chi_\omega\tilde{\Pi}_{\qv,\omega}$
also goes to a constant, so $\Lambda_\omega$ has to go to a constant as well.

The nonzero retarded interaction at high frequencies is notably different from EDMFT and DB with the ``dual'' self--consistency condition. 
This frequency independent retarded interaction can be considered as a change in the instantaneous effective impurity interaction. Physically, this is the screening of the impurity by the collective modes.

\section{Conclusion}

In this paper, we formulated the inherently self--consistent dual boson scheme, capable of treating electron structure and collective excitations
in correlated systems. The scheme employs the effective impurity problem (\ref{eq:s:imp}) with fermionic and bosonic hybridization functions.
We have shown that a proper choice of the retarded interaction leads to a significant simplification of the DB perturbation theory: all diagrams with local two--particle lines can be removed.
In this way, the effective impurity problem contains more information about the nonlocal bosonic (e.g. charge, spin) fluctuations.
Physically, the nonlocal bosonic fluctuations are determined \emph{both} by the direct boson--boson interaction in the Hamiltonian and by the collective behavior of the fermions. These two phenomena need to be taken into account on the same footing. 

We have found that the nonlocal charge fluctuations make the system more insulating. However, to make a truly meaningful statement on this issue, the nonlocal spin fluctuations should be taken into account on the same footing, since they are expected to have a compensating effect. 

Perhaps somewhat surprisingly, an impurity model with a static $U$ is not the optimal reference for the Hubbard model with local interaction only. In finite dimensions, nonlocal charge and spin fluctuations can have large effects~\cite{Schafer15} and ideally the impurity problem should know about this. DMFT overestimates the N\'{e}el temperature~\cite{Brener08,Otsuki14,Schafer15,Hirschmeier15} since the feedback of the strong antiferromagnetic fluctuations on the impurity model is not taken into account. In 2d, these long--range antiferromagnetic fluctuations should bring down the N\'{e}el temperature all the way to zero according to the Mermin--Wagner theorem.

In the presence of nearest--neighbor interaction $V$, we have found that the effective impurity interaction depends approximately linearly on $V$, as expected from the Peierls--Feynman--Bogoliubov variational principle~\cite{Schuler13}.

We have also presented a proof of the charge conservation law in DB, and have shown that charge conservation can be achieved independently of the choice of retarded interaction. Numerically the DB charge susceptibility indeed satisfies this law. Charge conservation is important for a proper description of long--wavelength modes.

Finally, it is worth to draw an analogy to the DMFT paradigm. DMFT is known to be the best single--impurity 
method describing one--particle physics of correlated systems in a fully self--consistent way. It is important that it uses only the single--particle Green's function of the 
impurity model. This is because of the proper choice of the fermionic hybridization function, which allows low--order dual fermion corrections to vanish (as these would otherwise include vertex parts of the impurity).
The theory presented above follows the same ideology for the collective excitations: the calculation requires knowledge of one-- and two--particle properties of the self--consistent impurity model, whereas the proper choice of the fermionic and bosonic hybridization functions eliminates the diagrams with higher--order vertexes.

\acknowledgments

M.I.K., E.A.S. and E.G.C.P. v. L. acknowledge support from ERC Advanced Grant 338957 FEMTO/NANO and from NWO via Spinoza Prize, A. I. L. from the DFG Research Unit FOR 1346 and
A. K. acknowledges support of FASO, Russian Federation (theme Electron No. 01201463326). A.K. and A.R. acknowledge Dynasty Foundation. E.A.S. also acknowledges the Russian Quantum Center for the hospitality and support by the EU Network FP7-PEOPLE-2013-IRSES Grant No 612624 ``InterNoM''.
A. R. is grateful to RFBR, Grant No. 14-02-01219 and A.A.K., A.I.L., and M.I.K. acknowledge support from Act 211 Government of the Russian Federation, Contract No. 02.A03.21.0006.
We employed a modified version of an open source implementation of the hybridization expansion quantum impurity solver~\cite{Hafermann13} with improved estimators~\cite{Hafermann14} for the vertices, based on the ALPS libraries~\cite{ALPS2}.

\appendix

\section{Sum rule for the fermion--boson vertex}
\label{app:sumrule}

The fermion--boson vertex $\lambda$ and the fermion--fermion vertex $\gamma$ are related~\cite{Rubtsov12,vanLoon14-2}, since they are both two--particle impurity correlators
 \begin{align}
 \lambda_{\nu\omega}^{\sigma} = \chi^{-1}_{\omega} \left(\frac{1}{\beta}\sum_{\sigma'\nu'}\gamma^{\sigma\sigma'}_{\nu\nu'\omega}g_{\nu'\sigma'}g_{\nu'+\omega\sigma'} -1\right).
 \end{align}
The impurity susceptibility is also a two--particle correlator, and it is related to the fermion--fermion vertex by a ladder equation~\cite{Hafermann14-2}
 \begin{align}
  \chi_\omega =
  \sum_{\nu\sigma} g_{\nu\sigma}g_{\nu+\omega,\sigma}
  -\sum_{\nu\nu'\sigma\sigma'}
  g_{\nu\sigma}g_{\nu+\omega,\sigma} \gamma_{\nu\nu'\omega} g_{\nu'\sigma}g_{\nu'+\omega,\sigma}.
 \end{align}
Combining these equations gives a useful sum rule for the fermion--boson vertex
 \begin{align}
  \sum_{\nu\sigma} g_{\nu\sigma}g_{\nu+\omega\sigma} \chi_\omega \lambda_{\nu\omega}^{\sigma} =& \sum_{\nu\nu'\sigma\sigma'} g_{\nu\sigma}g_{\nu+\omega,\sigma} \gamma_{\nu\nu'\omega} g_{\nu'\sigma}g_{\nu'+\omega,\sigma}
  \notag \\
  &- \sum_{\nu\sigma} g_{\nu\sigma}g_{\nu+\omega,\sigma} \notag \\
  =& - \chi_\omega \notag \\
  \sum_{\nu\sigma} g_{\nu\sigma}g_{\nu+\omega\sigma} \lambda_{\nu\omega}^{\sigma} =& -1.
  \label{eq:app:sumrulelambda}
 \end{align}

This identity can also be obtained straightforwardly from the definition of $\lambda$
\begin{align}
 \sum_{\nu\sigma} g_{\nu\sigma}g_{\nu+\omega\sigma} \chi_\omega \lambda_{\nu\omega}^{\sigma}
 =&
 -\sum_{\nu\sigma} \av{c_{\nu\sigma}c^*_{\nu+\omega\sigma} n_\omega} - g_{\nu\sigma}\av{n}\delta_{\omega} \notag \\
 =& \av{ n_\omega n_\omega} - \av{n}\av{n}\delta_\omega \notag\\
 =& -\chi_\omega.
\end{align}

\section{Impurity vertex functions}
\label{app:vertices}
The three--particle vertex functions are given by
\begin{align}
&\gamma^{6,0}_{1,2,3,4,5,6} =
(g_1g_2g_3g_4g_5g_6)^{-1}\times \notag \\
&\big[\av{c_1c_2c_3c^{*}_4c^{*}_5c^{*}_6} -
\av{c_1c^{*}_4}\gamma^{4,0}g_2g_3g_5g_6 + \notag \\
&\,\,\,\av{c_1c^{*}_5}\gamma^{4,0}g_2g_3g_4g_6 -
\av{c_1c^{*}_6}\gamma^{4,0}g_2g_3g_4g_5 + \notag \\
&\,\,\,\av{c_2c^{*}_4}\gamma^{4,0}g_1g_3g_5g_6 -
\av{c_2c^{*}_5}\gamma^{4,0}g_1g_3g_4g_6 + \notag \\
&\,\,\,\av{c_2c^{*}_6}\gamma^{4,0}g_1g_3g_4g_5 -
\av{c_3c^{*}_4}\gamma^{4,0}g_1g_2g_5g_6 + \notag \\
&\,\,\,\av{c_3c^{*}_5}\gamma^{4,0}g_1g_2g_4g_6 -
\av{c_3c^{*}_6}\gamma^{4,0}g_1g_2g_4g_5 - \notag \\
&\,\,\,\av{c_1c^{*}_4}\av{c_2c^{*}_6}\av{c_3c^{*}_5} +
\av{c_1c^{*}_4}\av{c_2c^{*}_5}\av{c_3c^{*}_6} - \notag \\
&\,\,\,\av{c_1c^{*}_5}\av{c_2c^{*}_4}\av{c_3c^{*}_6} +
\av{c_1c^{*}_5}\av{c_2c^{*}_6}\av{c_3c^{*}_4} - \notag \\
&\,\,\,\av{c_1c^{*}_6}\av{c_2c^{*}_5}\av{c_3c^{*}_4} +
\av{c_1c^{*}_6}\av{c_2c^{*}_4}\av{c_3c^{*}_5}\big]
\label{def:6.0}
\end{align}
whereas $\gamma^{4,1}$ and $\gamma^{2,2}$ can be expressed via $\gamma^{6,0}$ as follows:
\begin{align}
&\gamma^{4,1}_{1,2,5,6;3} = \chi_3^{-1}\sum\limits_{4}\big[
\gamma^{6,0}_{1,2,3+4,4,5,6}g_{3+4}g_{4} + \notag \\ &\delta_{1,4}\gamma^{4,0}_{2,3+4,5,6}g_{3+4} -
\delta_{3+4,5}\gamma^{4,0}_{1,2,4,6}g_{4} +
\delta_{3+4,6}\gamma^{4,0}_{1,2,4,5}g_{4}
\big]
\label{def:4.1}
\end{align}
\begin{align}
\gamma^{2,2}_{1,6;2,3} = (\chi_2\chi&_3)^{-1}\times \notag \\
\sum\limits_{4,5}\big[&
\gamma^{6,0}_{1,4+2,4,5+3,5,6}g_{4}g_{5}g_{4+2}g_{5+3} + \notag \\
&\delta_{1,5}\delta_{4+2,6}\av{c_{5+3}c^{*}_4} + \notag \\
&\delta_{1,4}\delta_{5+3,6}\av{c_{4+2}c^{*}_5} + \notag \\ &\delta_{1,4}\gamma^{4,0}_{5,4+2,5+3,6}g_{5}g_{4+2}g_{5+3} - \notag \\
&\delta_{1,5}\gamma^{4,0}_{4,6,4+2,5+3}g_{4}g_{6}g_{4+2}g_{5+3} - \notag \\
&\delta_{1,4}\gamma^{4,0}_{5,6,4+2,5+3}g_{5}g_{6}g_{4+2}g_{5+3} - \notag \\
&\delta_{6,5+3}\gamma^{4,0}_{1,4,5,4+2}g_{1}g_{4}g_{5}g_{4+2} + \notag \\
&\delta_{6,4+2}\gamma^{4,0}_{1,4,5,5+3}g_{1}g_{4}g_{5}g_{5+3}
\big].
\label{def:2.2}
\end{align}
This shows that $\gamma^{2,2}$ contains one--particle and two--particle reducible contributions, whereas $\gamma^{4,1}$ only contains the two--particle reducible part.

\section{Equivalence of the two self--consistency conditions within the EDMFT}
\label{equality}
In EDMFT, where the dual polarization is equal to zero, $\tilde{\Pi}_{\qv,\omega}=0$, the lattice susceptibility is given by \eqref{eq:xfrompid}
\begin{equation}
X_{q,\omega} = (\chi^{-1}_{\omega}+\Lambda_{\omega}-V_{q})^{-1}.
\end{equation}
Then, the ``lattice'' self--consistency condition is
\begin{equation}
\chi_{\omega} = \sum\limits_{q}(\chi^{-1}_{\omega}+\Lambda_{\omega}-V_{q})^{-1}.
\label{equal1}
\end{equation}
At the same time, since $\tilde{\Pi}_{\qv,\omega}=0$, the dual susceptibility $\tilde{X}$ is equal to the bare dual susceptibility \eqref{eq:barebosonpropagator}. The old self--consistency condition \eqref{eq:sc:db} is
\begin{equation}
0 = \sum\limits_{q}(\chi^{-1}_{\omega}+\Lambda_{\omega}-V_{q})^{-1} - \chi_{\omega},
\label{equal2}
\end{equation}
and \eqref{equal1} and \eqref{equal2} are clearly equivalent. This proves the equivalence of the self--consistency conditions \eqref{eq:sc:db} and \eqref{eq:sc:edmft} in EDMFT.

\section{Relation between lattice and dual susceptibilities}
\label{app:DX}

The self--consistency condition \eqref{eq:sc:edmft} is formulated in terms of lattice quantities. We wish to have an interpretation of this 
self--consistency condition in the dual perturbation theory, which requires us to rewrite the self--consistency condition in terms of dual quantities. 
To achieve this, we start by rewriting the bare dual bosonic propagator of Eq.~\eqref{eq:barebosonpropagator}
\begin{align}
\tilde{X}^{0} 
&=
\left[\chi^{-1}-(V-\Lambda)\right]^{-1}-\chi \notag\\
&=
\frac{\chi(V-\Lambda)\chi}{1-(V-\Lambda)\chi}.
\end{align}
The momentum and frequency labels in these expression have been suppressed to simplify the notation.
The Dyson equation gives the renormalized dual susceptibility
\begin{align}
\tilde{X}
&=
\frac{\tilde{X}^{0}}{1-\tilde{X}^{0}\tilde{\Pi}} \notag \\
&=
\frac{\chi(V-\Lambda)\chi}{1-(V-\Lambda)(\chi+\chi\tilde{\Pi}\chi)}.
\label{box1}
\end{align}
The lattice susceptibility [Eq.~\eqref{eq:xfrompid}] can be written in a similar way as
\begin{align}
X = \frac{\chi+\chi \tilde{\Pi}\chi}{1 - (V-\Lambda)(\chi+\chi \tilde{\Pi}\chi)}.
\label{box11}
\end{align}
This expression can be illustrated by the diagrammatic series
\begin{align}
\includegraphics[width=1\linewidth]{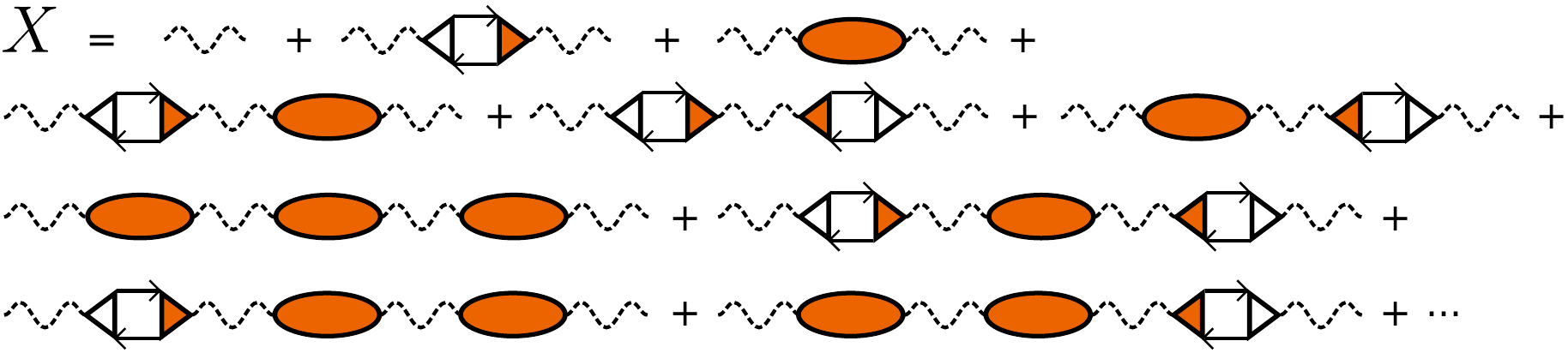}, \label{eq:Xdiagrams}
\end{align}
where the dashed line is the impurity susceptibility $\chi$, the filled ellipse denotes the interaction $V-\Lambda$ and the triangles with fermion lines are the dual polarization $\tilde{\Pi}$.

Using the Eq.~\eqref{box1} and Eq.~\eqref{box11}, one can get the relation between lattice and dual susceptibilities
\begin{align}
X = \tilde{X}\frac{\chi+\chi\tilde{\Pi}\chi}{\chi(V-\Lambda)\chi}.
\label{box2}
\end{align}
Finally, we wish to get rid of $V-\Lambda$. To do this we solve Eq.~\eqref{box1} for $V-\Lambda$ and substitute the result
\begin{align}
V-\Lambda = \frac{\tilde{X}}{\tilde{X}(\chi+\chi\tilde{\Pi}\chi) + \chi^2}.
\end{align}
in Eq.~\eqref{box2}. This results in Eq.~\eqref{scline}
\begin{align}
X &= \tilde{X}\frac{(\chi+\chi\tilde{\Pi}\chi)\big[\tilde{X}(\chi+\chi\tilde{\Pi}\chi) + \chi^2\big]}{\chi\tilde{X}\chi} \notag \\
&= (1+\chi\tilde{\Pi})\big[\tilde{X}(1+\tilde{\Pi}\chi) + \chi\big].
\end{align}

\section{One-- and two--particle reducible parts}
\label{app:12reduc}
Here, we study the reducible parts of the higher--order vertices with respect to the purely fermionic vertex functions $\gamma^{4,0}$ and $\gamma^{6,0}$.
We start with the simplest one--particle correction to the self energy,
from the reducible part of Fig.~\ref{Transformation}c), where the one--particle reducible contribution comes from \eqref{def:2.2}.
Just like $\gamma^{2,1}=\lambda$, $\gamma^{2,2}$ is nonzero even in noninteracting systems.
Diagrammatically, the associated dual self--energy is
\begin{align}
\tilde{\Sigma}^{c}_{1PR} = 
\includegraphics[width=0.62\linewidth]{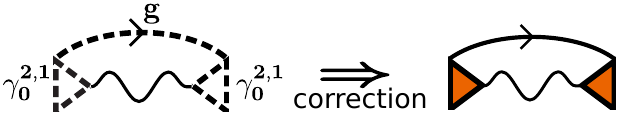}.
\label{eq:S1pr}
\end{align}
where the dashed triangular vertex is the bare fermion--boson vertex $\gamma^{2,1}_0=-1/\chi$. This one--particle reducible part of $\Sigma^{c}$ acts as a correction to the dual self--energy in Eq.~\ref{eq:sigma}.

The dual polarization for this case also has a simple form,
\begin{align}
\tilde{\Pi}^{c}=
\includegraphics[width=0.13\linewidth]{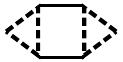}.
\label{eq:Pi1pt}
\end{align}
This contribution to the polarization function corresponds to the limit $U = \Lambda = 0$. In that case, the interaction $V$ is small and we can
expand the relation for the dual susceptibility to first order in $V$
\begin{align}
\tilde{X} = \frac{\chi(V-\Lambda)\chi}{1-(V-\Lambda)(\chi+\chi\tilde{\Pi}\chi)} = \chi{}V\chi.
\end{align}
Since the local part of interaction $V_{q}$ is equal to zero, the local part of the dual susceptibility is equal to zero as well.
This means that the self--energy $\tilde{\Sigma}^{c}_{1PR}$ in \eqref{eq:S1pr} is equal to zero and therefore only bare fermionic lines remain in polarization function \eqref{eq:Pi1pt}.

In this case the lattice susceptibility is~\cite{vanLoon14-2}
\begin{align}
X = \frac{GG}{1-VGG}.
\end{align}
This result is exactly the RPA relation for the susceptibility.
It means that for the simplest case the self--consistent DB approach reproduces the result of the Random Phase Approximation (RPA) method. In this limit of weak interaction, RPA indeed correctly describes the bosonic degrees of freedom~\cite{Pines66} and self--consistent dual boson reduces to RPA in the same limit.

Now, we can go further and study the two--particle contribution from the reducible parts of the vertices $\gamma^{4,1}$ and $\gamma^{2,2}$, see Eqs.~\eqref{def:4.1} and \eqref{def:2.2}. 
Including the four--point vertex $\gamma^{4,0}$, the additional reducible parts of the diagrams for the self--energy are
\begin{align}
\tilde{\Sigma}^{c}_{2PR} = 
\includegraphics[width=0.72\linewidth]{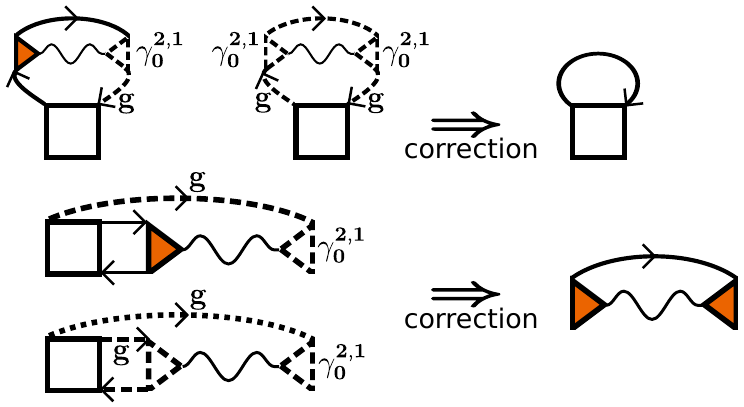}
\end{align}
When the impurity Green's function $g_\nu$ is small, e.g., at high
temperature and large U, these terms become small corrections. In the
limit of small U, they also become small corrections since the
three--particle correlators are small themselves.
Thus, the main contribution of the three--particle correlation interaction functions to the self--energy comes from irreducible parts of $\gamma^{4,1}$ and $\gamma^{2,2}$ and the vertex function $\gamma^{6,0}$ itself.

\section{Contribution of higher--order vertices to the self--energy}
\label{app:small_red}

A main advantage of the DB approach is the fact that it can be successfully applied to both the large and small $U$ limits. For the small $U$ case the diagrammatic expansion has a perturbative form and the higher--order vertices in $U$ can be neglected.
Here we analyze the physical irreducible contribution from the higher--order vertices, which are similar to those removed by the bosonic self--consistency condition. Other reducible contributions are also small for this case, as was shown in Appendix~\ref{app:12reduc}.

The irreducible part of $\gamma^{2,2}$ vertex function can be presented in terms of the six--point vertex function as
\begin{align}
\gamma^{2,2}_{1,6;2,3} = (\chi_2\chi_3)^{-1}\sum\limits_{4,5}
\gamma^{6,0}_{1,4,4+2,5,5+3,6} g_{4}g_{5}g_{4+2}g_{5+3}.
\end{align}
One can rewrite this equation in following way
\begin{align}
&\gamma^{2,2}_{1,6;2,3}\sum_{4,5}g_{4}g_{4+2}g_{5}g_{5+3}\big(1 + \sum_{4'}\gamma^{4,0}_{4,4+2,4',4'+2}g_{4'}g_{4'+2}\big)\times \notag \\
&\big(1 + \sum_{5'}\gamma^{4,0}g_{5'}g_{5'+3}\big) =  \sum\limits_{4,5}
\gamma^{6,0}_{1,4,4+2,5,5+3,6} g_{4}g_{4+2}g_{5}g_{5+3},
\end{align}
or equally
\begin{align}
&\sum_{4,5}\gamma^{2,2}_{1,6;2,3}\chi_2\chi_3 g_{4}g_{4+2}g_{5}g_{5+3}\times \notag \\
&\big(-\frac{1}{\chi_2} - \frac{1}{\chi_2}\sum_{4'}\gamma^{4,0}_{4,4+2,4',4'+2}g_{4'}g_{4'+2}\big)\times \notag \\
&\big(-\frac{1}{\chi_3} - \frac{1}{\chi_3}\sum_{5'}\gamma^{4,0}_{5,5+3,5',5'+3}g_{5'}g_{5'+3}\big)= \notag \\
&\sum\limits_{4,5}\gamma^{6,0}_{1,4,4+2,5,5+3,6} g_{4}g_{4+2}g_{5}g_{5+3}.
\end{align}
Using the relation \eqref{eq:lambda} we get
\begin{align}
&\sum_{4,5}\gamma^{2,2}_{1,6;2,3}\chi_2\chi_3g_{4}g_{4+2}g_{5}g_{5+3}\times \notag \\
&\big(-\frac{2}{\chi_2} - \gamma^{2,1}_{4,4+2;2}\big)\big(-\frac{2}{\chi_3} - \gamma^{2,1}_{5,5+3;3}\big) = \notag \\
&\sum\limits_{4,5}\gamma^{6,0}_{1,4,4+2,5,5+3,6}g_{4}g_{4+2}g_{5}g_{5+3}.
\label{eq:neglecting}
\end{align}
If we look again on the relation \eqref{eq:lambda} we can realize that the bare vertex function $-\frac{1}{\chi}$ can be expressed through
the local three-- and four--point vertex functions and local Green's function:
\begin{align}
\includegraphics[width=0.5\linewidth]{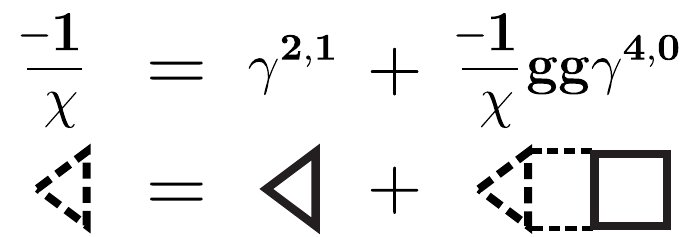}
\end{align}
The contribution of the six--point vertex to the self--energy $\tilde{\Sigma}^{b}$ now looks as follows:
\begin{align}
\tilde{\Sigma}^{b}=
\includegraphics[width=0.25\linewidth]{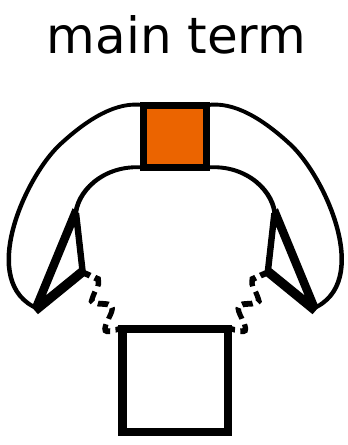}
+
\includegraphics[width=0.25\linewidth]{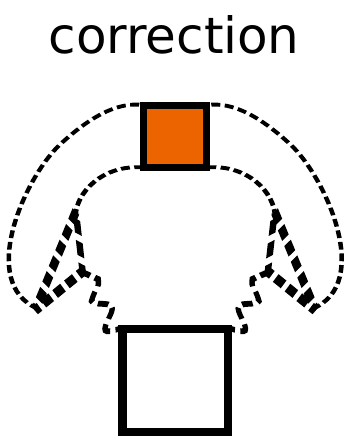}.
\end{align}
When the two--particle vertex or impurity Green function $g$ are small, 
then all terms with the dashed lines are accounted for the small correction. Therefore we consider only the main term in the
bare three--point vertex $-\frac{1}{\chi_2}\rightarrow\gamma^{2,1}_{4,4+2;2}$ with the proper summation over the fermionic frequency.
This is the case for example when the three--point vertex weakly depends on the fermionic frequency.
  Then one can obtain the following relation
\begin{align}
&\sum\limits_{4,5}\gamma^{6,0}_{1,4,4+2,5,5+3,6} g_{4}g_{4+2}g_{5}g_{5+3}= \notag \\
&\sum_{4,5}\gamma^{2,2}_{1,6;2,3}\chi_2\chi_3\gamma^{2,1}_{4,4+2;2}\gamma^{2,1}_{5,5+3;3} g_{4}g_{4+2}g_{5}g_{5+3}. \label{F6}
\end{align}
Similarly to the six--point case we can obtain the relation for the five--point vertex as follows
\begin{align}
\sum\limits_{4}\gamma^{4,1}_{1,4,4+2,6;3} g_{4}g_{4+2}=
\sum_{4}\gamma^{2,2}_{1,6;2,3}\chi_2\gamma^{2,1}_{4,4+2;3} g_{4}g_{4+2}, \label{F7}
\end{align}
The expressions (\ref{F6}) and (\ref{F7}) are similar to those shown in Fig.~\ref{Transformation} and differ from them
only by the replacement of the local Green's functions by their dual counterparts. One can expect, however, that
these diagrams, being physically important, indeed give the dominant contribution to the self--energy, which therefore can be cancelled by the self--consistency condition, as it was shown in \eqref{sl0}.

\section{Quadratic dependence on $V$ in EDMFT}

\label{app:edmft:fail}
Our EDMFT results show a quadratic dependence on $V$.
This can be understood from a symmetry of the square lattice, and of hypercubic lattices in general. The Fourier transform of the nearest--neighbor interaction is
\begin{align}
 V_\qv = 2V \sum_{\alpha=1}^d \cos(k_\alpha).
\end{align}
All the momenta occur in pairs $\qv,\qv'$, with $\qv'=\qv+(\pi,\pi)$ and $V_\qv = - V_{\qv'}$.
In the EDMFT formalism, $V_\qv$ is the only quantity carrying momentum dependence, and it only occurs in formulas with a momentum--averaging, such as the self--consistency condition. Changing the sign of the interaction, i.e., $V'_\qv = - V_{\qv}$ does not change the self--consistency condition:
\begin{align}
 \chi_\omega =&
 \sum_{\qv} \frac{1}{\chi_\omega^{-1}+\Lambda_\omega - V_{\qv}}
= \sum_{\qv} \frac{1}{\chi_\omega^{-1}+\Lambda_\omega - [-V_{\qv'}] }
 \notag \\
=& \sum_{\qv'} \frac{1}{\chi_\omega^{-1}+\Lambda_\omega - [-V_{\qv'}]}
= \sum_{\qv'} \frac{1}{\chi_\omega^{-1}+\Lambda_\omega - V'_{\qv'}}.
\end{align}
This shows that the EDMFT result in a hypercubic lattice does not depend on the sign of the nearest--neighbor interaction. This is not what is expected physically, attractive and repulsive nearest--neighbor interactions are notably different.

In self--consistent DB the results \emph{do} depend on the sign of the interaction. Since there is an additional quantity with momentum dependence, $\tilde{\Pi}_{\qv\omega}$, the sign flipping argument of the previous paragraph no longer applies.

Diagrammatically, this can be seen from Eq.~\eqref{eq:Xdiagrams}, where the momentum dependence comes in through the filled ellipse (via the interaction $V_\qv$) and through the fermionic bubbles (the dual polarization). 
EDMFT does not contain the latter, so all of the momentum dependence comes from $V_\qv$. 
Due to the symmetry of the square lattice, changing the nearest-neighbour interaction from repulsive to attractive only rearranges the interaction terms and leaves their momentum averages the same.
In dual boson, Eq.~\eqref{eq:Xdiagrams} contains contributions where the both the dual polarization and the interaction appear. Exactly these diagrams give the contributions linear in $V$.

\bibliography{main}

\begin{thebibliography}{62}%
\makeatletter
\providecommand \@ifxundefined [1]{%
 \@ifx{#1\undefined}
}%
\providecommand \@ifnum [1]{%
 \ifnum #1\expandafter \@firstoftwo
 \else \expandafter \@secondoftwo
 \fi
}%
\providecommand \@ifx [1]{%
 \ifx #1\expandafter \@firstoftwo
 \else \expandafter \@secondoftwo
 \fi
}%
\providecommand \natexlab [1]{#1}%
\providecommand \enquote  [1]{``#1''}%
\providecommand \bibnamefont  [1]{#1}%
\providecommand \bibfnamefont [1]{#1}%
\providecommand \citenamefont [1]{#1}%
\providecommand \href@noop [0]{\@secondoftwo}%
\providecommand \href [0]{\begingroup \@sanitize@url \@href}%
\providecommand \@href[1]{\@@startlink{#1}\@@href}%
\providecommand \@@href[1]{\endgroup#1\@@endlink}%
\providecommand \@sanitize@url [0]{\catcode `\\12\catcode `\$12\catcode
  `\&12\catcode `\#12\catcode `\^12\catcode `\_12\catcode `\%12\relax}%
\providecommand \@@startlink[1]{}%
\providecommand \@@endlink[0]{}%
\providecommand \url  [0]{\begingroup\@sanitize@url \@url }%
\providecommand \@url [1]{\endgroup\@href {#1}{\urlprefix }}%
\providecommand \urlprefix  [0]{URL }%
\providecommand \Eprint [0]{\href }%
\providecommand \doibase [0]{http://dx.doi.org/}%
\providecommand \selectlanguage [0]{\@gobble}%
\providecommand \bibinfo  [0]{\@secondoftwo}%
\providecommand \bibfield  [0]{\@secondoftwo}%
\providecommand \translation [1]{[#1]}%
\providecommand \BibitemOpen [0]{}%
\providecommand \bibitemStop [0]{}%
\providecommand \bibitemNoStop [0]{.\EOS\space}%
\providecommand \EOS [0]{\spacefactor3000\relax}%
\providecommand \BibitemShut  [1]{\csname bibitem#1\endcsname}%
\let\auto@bib@innerbib\@empty
\bibitem [{\citenamefont {Tosatti}\ and\ \citenamefont
  {Anderson}(1974)}]{Tosatti74}%
  \BibitemOpen
  \bibfield  {author} {\bibinfo {author} {\bibfnamefont {E.}~\bibnamefont
  {Tosatti}}\ and\ \bibinfo {author} {\bibfnamefont {P.~W.}\ \bibnamefont
  {Anderson}},\ }\href {http://stacks.iop.org/1347-4065/13/i=S2/a=381}
  {\bibfield  {journal} {\bibinfo  {journal} {Japanese Journal of Applied
  Physics}\ }\textbf {\bibinfo {volume} {13}},\ \bibinfo {pages} {381}
  (\bibinfo {year} {1974})}\BibitemShut {NoStop}%
\bibitem [{\citenamefont {Hansmann}\ \emph {et~al.}(2013)\citenamefont
  {Hansmann}, \citenamefont {Ayral}, \citenamefont {Vaugier}, \citenamefont
  {Werner},\ and\ \citenamefont {Biermann}}]{Hansmann13}%
  \BibitemOpen
  \bibfield  {author} {\bibinfo {author} {\bibfnamefont {P.}~\bibnamefont
  {Hansmann}}, \bibinfo {author} {\bibfnamefont {T.}~\bibnamefont {Ayral}},
  \bibinfo {author} {\bibfnamefont {L.}~\bibnamefont {Vaugier}}, \bibinfo
  {author} {\bibfnamefont {P.}~\bibnamefont {Werner}}, \ and\ \bibinfo {author}
  {\bibfnamefont {S.}~\bibnamefont {Biermann}},\ }\href {\doibase
  10.1103/PhysRevLett.110.166401} {\bibfield  {journal} {\bibinfo  {journal}
  {Phys. Rev. Lett.}\ }\textbf {\bibinfo {volume} {110}},\ \bibinfo {pages}
  {166401} (\bibinfo {year} {2013})}\BibitemShut {NoStop}%
\bibitem [{\citenamefont {Huang}\ \emph {et~al.}(2014)\citenamefont {Huang},
  \citenamefont {Ayral}, \citenamefont {Biermann},\ and\ \citenamefont
  {Werner}}]{Huang14}%
  \BibitemOpen
  \bibfield  {author} {\bibinfo {author} {\bibfnamefont {L.}~\bibnamefont
  {Huang}}, \bibinfo {author} {\bibfnamefont {T.}~\bibnamefont {Ayral}},
  \bibinfo {author} {\bibfnamefont {S.}~\bibnamefont {Biermann}}, \ and\
  \bibinfo {author} {\bibfnamefont {P.}~\bibnamefont {Werner}},\ }\href
  {\doibase 10.1103/PhysRevB.90.195114} {\bibfield  {journal} {\bibinfo
  {journal} {Phys. Rev. B}\ }\textbf {\bibinfo {volume} {90}},\ \bibinfo
  {pages} {195114} (\bibinfo {year} {2014})}\BibitemShut {NoStop}%
\bibitem [{\citenamefont {van Loon}\ \emph
  {et~al.}(2014{\natexlab{a}})\citenamefont {van Loon}, \citenamefont
  {Lichtenstein}, \citenamefont {Katsnelson}, \citenamefont {Parcollet},\ and\
  \citenamefont {Hafermann}}]{vanLoon14-2}%
  \BibitemOpen
  \bibfield  {author} {\bibinfo {author} {\bibfnamefont {E.~G. C.~P.}\
  \bibnamefont {van Loon}}, \bibinfo {author} {\bibfnamefont {A.~I.}\
  \bibnamefont {Lichtenstein}}, \bibinfo {author} {\bibfnamefont {M.~I.}\
  \bibnamefont {Katsnelson}}, \bibinfo {author} {\bibfnamefont
  {O.}~\bibnamefont {Parcollet}}, \ and\ \bibinfo {author} {\bibfnamefont
  {H.}~\bibnamefont {Hafermann}},\ }\href {\doibase 10.1103/PhysRevB.90.235135}
  {\bibfield  {journal} {\bibinfo  {journal} {Phys. Rev. B}\ }\textbf {\bibinfo
  {volume} {90}},\ \bibinfo {pages} {235135} (\bibinfo {year}
  {2014}{\natexlab{a}})}\BibitemShut {NoStop}%
\bibitem [{\citenamefont {Mott}(1974)}]{Mott74}%
  \BibitemOpen
  \bibfield  {author} {\bibinfo {author} {\bibfnamefont {N.}~\bibnamefont
  {Mott}},\ }\href@noop {} {\emph {\bibinfo {title} {Metal-Insulator
  Transitions}}}\ (\bibinfo  {publisher} {Taylor \& Francis},\ \bibinfo {year}
  {1974})\BibitemShut {NoStop}%
\bibitem [{\citenamefont {Imada}\ \emph {et~al.}(1998)\citenamefont {Imada},
  \citenamefont {Fujimori},\ and\ \citenamefont {Tokura}}]{Imada98}%
  \BibitemOpen
  \bibfield  {author} {\bibinfo {author} {\bibfnamefont {M.}~\bibnamefont
  {Imada}}, \bibinfo {author} {\bibfnamefont {A.}~\bibnamefont {Fujimori}}, \
  and\ \bibinfo {author} {\bibfnamefont {Y.}~\bibnamefont {Tokura}},\ }\href
  {\doibase 10.1103/RevModPhys.70.1039} {\bibfield  {journal} {\bibinfo
  {journal} {Rev. Mod. Phys.}\ }\textbf {\bibinfo {volume} {70}},\ \bibinfo
  {pages} {1039} (\bibinfo {year} {1998})}\BibitemShut {NoStop}%
\bibitem [{\citenamefont {Walz}(2002)}]{Walz02}%
  \BibitemOpen
  \bibfield  {author} {\bibinfo {author} {\bibfnamefont {F.}~\bibnamefont
  {Walz}},\ }\href {http://stacks.iop.org/0953-8984/14/i=12/a=203} {\bibfield
  {journal} {\bibinfo  {journal} {Journal of Physics: Condensed Matter}\
  }\textbf {\bibinfo {volume} {14}},\ \bibinfo {pages} {R285} (\bibinfo {year}
  {2002})}\BibitemShut {NoStop}%
\bibitem [{\citenamefont {van Loon}\ \emph
  {et~al.}(2014{\natexlab{b}})\citenamefont {van Loon}, \citenamefont
  {Hafermann}, \citenamefont {Lichtenstein}, \citenamefont {Rubtsov},\ and\
  \citenamefont {Katsnelson}}]{vanLoon14}%
  \BibitemOpen
  \bibfield  {author} {\bibinfo {author} {\bibfnamefont {E.~G. C.~P.}\
  \bibnamefont {van Loon}}, \bibinfo {author} {\bibfnamefont {H.}~\bibnamefont
  {Hafermann}}, \bibinfo {author} {\bibfnamefont {A.~I.}\ \bibnamefont
  {Lichtenstein}}, \bibinfo {author} {\bibfnamefont {A.~N.}\ \bibnamefont
  {Rubtsov}}, \ and\ \bibinfo {author} {\bibfnamefont {M.~I.}\ \bibnamefont
  {Katsnelson}},\ }\href {\doibase 10.1103/PhysRevLett.113.246407} {\bibfield
  {journal} {\bibinfo  {journal} {Phys. Rev. Lett.}\ }\textbf {\bibinfo
  {volume} {113}},\ \bibinfo {pages} {246407} (\bibinfo {year}
  {2014}{\natexlab{b}})}\BibitemShut {NoStop}%
\bibitem [{\citenamefont {Hafermann}\ \emph {et~al.}(2014)\citenamefont
  {Hafermann}, \citenamefont {van Loon}, \citenamefont {Katsnelson},
  \citenamefont {Lichtenstein},\ and\ \citenamefont
  {Parcollet}}]{Hafermann14-2}%
  \BibitemOpen
  \bibfield  {author} {\bibinfo {author} {\bibfnamefont {H.}~\bibnamefont
  {Hafermann}}, \bibinfo {author} {\bibfnamefont {E.~G. C.~P.}\ \bibnamefont
  {van Loon}}, \bibinfo {author} {\bibfnamefont {M.~I.}\ \bibnamefont
  {Katsnelson}}, \bibinfo {author} {\bibfnamefont {A.~I.}\ \bibnamefont
  {Lichtenstein}}, \ and\ \bibinfo {author} {\bibfnamefont {O.}~\bibnamefont
  {Parcollet}},\ }\href {\doibase 10.1103/PhysRevB.90.235105} {\bibfield
  {journal} {\bibinfo  {journal} {Phys. Rev. B}\ }\textbf {\bibinfo {volume}
  {90}},\ \bibinfo {pages} {235105} (\bibinfo {year} {2014})}\BibitemShut
  {NoStop}%
\bibitem [{\citenamefont {Moriya}(1985)}]{Moriya1985}%
  \BibitemOpen
  \bibfield  {author} {\bibinfo {author} {\bibfnamefont {T.}~\bibnamefont
  {Moriya}},\ }\href@noop {} {\emph {\bibinfo {title} {Spin fluctuations in
  itinerant electron magnetism}}}\ (\bibinfo  {publisher} {Springer-Verlag
  Berlin},\ \bibinfo {year} {1985})\BibitemShut {NoStop}%
\bibitem [{\citenamefont {Irkhin}\ and\ \citenamefont
  {Katsnelson}(1985)}]{Irkhin85}%
  \BibitemOpen
  \bibfield  {author} {\bibinfo {author} {\bibfnamefont {V.~Y.}\ \bibnamefont
  {Irkhin}}\ and\ \bibinfo {author} {\bibfnamefont {M.~I.}\ \bibnamefont
  {Katsnelson}},\ }\href {http://stacks.iop.org/0022-3719/18/i=21/a=013}
  {\bibfield  {journal} {\bibinfo  {journal} {Journal of Physics C: Solid State
  Physics}\ }\textbf {\bibinfo {volume} {18}},\ \bibinfo {pages} {4173}
  (\bibinfo {year} {1985})}\BibitemShut {NoStop}%
\bibitem [{\citenamefont {Secchi}\ \emph {et~al.}(2013)\citenamefont {Secchi},
  \citenamefont {Brener}, \citenamefont {Lichtenstein},\ and\ \citenamefont
  {Katsnelson}}]{Secchi13}%
  \BibitemOpen
  \bibfield  {author} {\bibinfo {author} {\bibfnamefont {A.}~\bibnamefont
  {Secchi}}, \bibinfo {author} {\bibfnamefont {S.}~\bibnamefont {Brener}},
  \bibinfo {author} {\bibfnamefont {A.}~\bibnamefont {Lichtenstein}}, \ and\
  \bibinfo {author} {\bibfnamefont {M.}~\bibnamefont {Katsnelson}},\ }\href
  {\doibase http://dx.doi.org/10.1016/j.aop.2013.03.006} {\bibfield  {journal}
  {\bibinfo  {journal} {Annals of Physics}\ }\textbf {\bibinfo {volume}
  {333}},\ \bibinfo {pages} {221 } (\bibinfo {year} {2013})}\BibitemShut
  {NoStop}%
\bibitem [{\citenamefont {Rubtsov}\ \emph {et~al.}(2009)\citenamefont
  {Rubtsov}, \citenamefont {Katsnelson}, \citenamefont {Lichtenstein},\ and\
  \citenamefont {Georges}}]{Rubtsov09}%
  \BibitemOpen
  \bibfield  {author} {\bibinfo {author} {\bibfnamefont {A.~N.}\ \bibnamefont
  {Rubtsov}}, \bibinfo {author} {\bibfnamefont {M.~I.}\ \bibnamefont
  {Katsnelson}}, \bibinfo {author} {\bibfnamefont {A.~I.}\ \bibnamefont
  {Lichtenstein}}, \ and\ \bibinfo {author} {\bibfnamefont {A.}~\bibnamefont
  {Georges}},\ }\href {\doibase 10.1103/PhysRevB.79.045133} {\bibfield
  {journal} {\bibinfo  {journal} {Phys. Rev. B}\ }\textbf {\bibinfo {volume}
  {79}},\ \bibinfo {pages} {045133} (\bibinfo {year} {2009})}\BibitemShut
  {NoStop}%
\bibitem [{\citenamefont {Rohringer}\ \emph {et~al.}(2011)\citenamefont
  {Rohringer}, \citenamefont {Toschi}, \citenamefont {Katanin},\ and\
  \citenamefont {Held}}]{Rohringer11}%
  \BibitemOpen
  \bibfield  {author} {\bibinfo {author} {\bibfnamefont {G.}~\bibnamefont
  {Rohringer}}, \bibinfo {author} {\bibfnamefont {A.}~\bibnamefont {Toschi}},
  \bibinfo {author} {\bibfnamefont {A.}~\bibnamefont {Katanin}}, \ and\
  \bibinfo {author} {\bibfnamefont {K.}~\bibnamefont {Held}},\ }\href {\doibase
  10.1103/PhysRevLett.107.256402} {\bibfield  {journal} {\bibinfo  {journal}
  {Phys. Rev. Lett.}\ }\textbf {\bibinfo {volume} {107}},\ \bibinfo {pages}
  {256402} (\bibinfo {year} {2011})}\BibitemShut {NoStop}%
\bibitem [{\citenamefont {Sch\"afer}\ \emph {et~al.}(2015)\citenamefont
  {Sch\"afer}, \citenamefont {Geles}, \citenamefont {Rost}, \citenamefont
  {Rohringer}, \citenamefont {Arrigoni}, \citenamefont {Held}, \citenamefont
  {Bl\"umer}, \citenamefont {Aichhorn},\ and\ \citenamefont
  {Toschi}}]{Schafer15}%
  \BibitemOpen
  \bibfield  {author} {\bibinfo {author} {\bibfnamefont {T.}~\bibnamefont
  {Sch\"afer}}, \bibinfo {author} {\bibfnamefont {F.}~\bibnamefont {Geles}},
  \bibinfo {author} {\bibfnamefont {D.}~\bibnamefont {Rost}}, \bibinfo {author}
  {\bibfnamefont {G.}~\bibnamefont {Rohringer}}, \bibinfo {author}
  {\bibfnamefont {E.}~\bibnamefont {Arrigoni}}, \bibinfo {author}
  {\bibfnamefont {K.}~\bibnamefont {Held}}, \bibinfo {author} {\bibfnamefont
  {N.}~\bibnamefont {Bl\"umer}}, \bibinfo {author} {\bibfnamefont
  {M.}~\bibnamefont {Aichhorn}}, \ and\ \bibinfo {author} {\bibfnamefont
  {A.}~\bibnamefont {Toschi}},\ }\href {\doibase 10.1103/PhysRevB.91.125109}
  {\bibfield  {journal} {\bibinfo  {journal} {Phys. Rev. B}\ }\textbf {\bibinfo
  {volume} {91}},\ \bibinfo {pages} {125109} (\bibinfo {year}
  {2015})}\BibitemShut {NoStop}%
\bibitem [{\citenamefont {Bloch}\ \emph {et~al.}(2012)\citenamefont {Bloch},
  \citenamefont {Dalibard},\ and\ \citenamefont {Nascimb{\`e}ne}}]{Bloch12}%
  \BibitemOpen
  \bibfield  {author} {\bibinfo {author} {\bibfnamefont {I.}~\bibnamefont
  {Bloch}}, \bibinfo {author} {\bibfnamefont {J.}~\bibnamefont {Dalibard}}, \
  and\ \bibinfo {author} {\bibfnamefont {S.}~\bibnamefont {Nascimb{\`e}ne}},\
  }\href@noop {} {\bibfield  {journal} {\bibinfo  {journal} {Nature Physics}\
  }\textbf {\bibinfo {volume} {8}},\ \bibinfo {pages} {267} (\bibinfo {year}
  {2012})}\BibitemShut {NoStop}%
\bibitem [{\citenamefont {Lewenstein}\ \emph {et~al.}(2012)\citenamefont
  {Lewenstein}, \citenamefont {Sanpera},\ and\ \citenamefont
  {Ahufinger}}]{Lewenstein12}%
  \BibitemOpen
  \bibfield  {author} {\bibinfo {author} {\bibfnamefont {M.}~\bibnamefont
  {Lewenstein}}, \bibinfo {author} {\bibfnamefont {A.}~\bibnamefont {Sanpera}},
  \ and\ \bibinfo {author} {\bibfnamefont {V.}~\bibnamefont {Ahufinger}},\
  }\href@noop {} {\emph {\bibinfo {title} {Ultracold Atoms in Optical Lattices:
  Simulating quantum many-body systems}}}\ (\bibinfo  {publisher} {Oxford
  University Press},\ \bibinfo {year} {2012})\BibitemShut {NoStop}%
\bibitem [{\citenamefont {van Loon}\ \emph
  {et~al.}(2015{\natexlab{a}})\citenamefont {van Loon}, \citenamefont
  {Katsnelson},\ and\ \citenamefont {Lemeshko}}]{vanLoon15-2}%
  \BibitemOpen
  \bibfield  {author} {\bibinfo {author} {\bibfnamefont {E.~G. C.~P.}\
  \bibnamefont {van Loon}}, \bibinfo {author} {\bibfnamefont {M.~I.}\
  \bibnamefont {Katsnelson}}, \ and\ \bibinfo {author} {\bibfnamefont
  {M.}~\bibnamefont {Lemeshko}},\ }\href {\doibase 10.1103/PhysRevB.92.081106}
  {\bibfield  {journal} {\bibinfo  {journal} {Phys. Rev. B}\ }\textbf {\bibinfo
  {volume} {92}},\ \bibinfo {pages} {081106} (\bibinfo {year}
  {2015}{\natexlab{a}})}\BibitemShut {NoStop}%
\bibitem [{\citenamefont {Wehling}\ \emph {et~al.}(2011)\citenamefont
  {Wehling}, \citenamefont {\ifmmode \mbox{\c{S}}\else \c{S}\fi{}a\ifmmode
  \mbox{\c{s}}\else \c{s}\fi{}\ifmmode \imath \else \i
  \fi{}o\ifmmode~\breve{g}\else \u{g}\fi{}lu}, \citenamefont {Friedrich},
  \citenamefont {Lichtenstein}, \citenamefont {Katsnelson},\ and\ \citenamefont
  {Bl\"ugel}}]{Wehling11}%
  \BibitemOpen
  \bibfield  {author} {\bibinfo {author} {\bibfnamefont {T.~O.}\ \bibnamefont
  {Wehling}}, \bibinfo {author} {\bibfnamefont {E.}~\bibnamefont {\ifmmode
  \mbox{\c{S}}\else \c{S}\fi{}a\ifmmode \mbox{\c{s}}\else \c{s}\fi{}\ifmmode
  \imath \else \i \fi{}o\ifmmode~\breve{g}\else \u{g}\fi{}lu}}, \bibinfo
  {author} {\bibfnamefont {C.}~\bibnamefont {Friedrich}}, \bibinfo {author}
  {\bibfnamefont {A.~I.}\ \bibnamefont {Lichtenstein}}, \bibinfo {author}
  {\bibfnamefont {M.~I.}\ \bibnamefont {Katsnelson}}, \ and\ \bibinfo {author}
  {\bibfnamefont {S.}~\bibnamefont {Bl\"ugel}},\ }\href {\doibase
  10.1103/PhysRevLett.106.236805} {\bibfield  {journal} {\bibinfo  {journal}
  {Phys. Rev. Lett.}\ }\textbf {\bibinfo {volume} {106}},\ \bibinfo {pages}
  {236805} (\bibinfo {year} {2011})}\BibitemShut {NoStop}%
\bibitem [{\citenamefont {Metzner}\ and\ \citenamefont
  {Vollhardt}(1989)}]{Metzner89}%
  \BibitemOpen
  \bibfield  {author} {\bibinfo {author} {\bibfnamefont {W.}~\bibnamefont
  {Metzner}}\ and\ \bibinfo {author} {\bibfnamefont {D.}~\bibnamefont
  {Vollhardt}},\ }\href {\doibase 10.1103/PhysRevLett.62.324} {\bibfield
  {journal} {\bibinfo  {journal} {Phys. Rev. Lett.}\ }\textbf {\bibinfo
  {volume} {62}},\ \bibinfo {pages} {324} (\bibinfo {year} {1989})}\BibitemShut
  {NoStop}%
\bibitem [{\citenamefont {Georges}\ \emph {et~al.}(1996)\citenamefont
  {Georges}, \citenamefont {Kotliar}, \citenamefont {Krauth},\ and\
  \citenamefont {Rozenberg}}]{Georges96}%
  \BibitemOpen
  \bibfield  {author} {\bibinfo {author} {\bibfnamefont {A.}~\bibnamefont
  {Georges}}, \bibinfo {author} {\bibfnamefont {G.}~\bibnamefont {Kotliar}},
  \bibinfo {author} {\bibfnamefont {W.}~\bibnamefont {Krauth}}, \ and\ \bibinfo
  {author} {\bibfnamefont {M.~J.}\ \bibnamefont {Rozenberg}},\ }\href {\doibase
  10.1103/RevModPhys.68.13} {\bibfield  {journal} {\bibinfo  {journal} {Rev.
  Mod. Phys.}\ }\textbf {\bibinfo {volume} {68}},\ \bibinfo {pages} {13}
  (\bibinfo {year} {1996})}\BibitemShut {NoStop}%
\bibitem [{\citenamefont {Hubbard}(1963)}]{Hubbard63}%
  \BibitemOpen
  \bibfield  {author} {\bibinfo {author} {\bibfnamefont {J.}~\bibnamefont
  {Hubbard}},\ }\href {\doibase 10.1098/rspa.1963.0204} {\bibfield  {journal}
  {\bibinfo  {journal} {Proceedings of the Royal Society of London. Series A.
  Mathematical and Physical Sciences}\ }\textbf {\bibinfo {volume} {276}},\
  \bibinfo {pages} {238} (\bibinfo {year} {1963})}\BibitemShut {NoStop}%
\bibitem [{\citenamefont {Hubbard}(1964)}]{Hubbard64}%
  \BibitemOpen
  \bibfield  {author} {\bibinfo {author} {\bibfnamefont {J.}~\bibnamefont
  {Hubbard}},\ }\href {\doibase 10.1098/rspa.1964.0190} {\bibfield  {journal}
  {\bibinfo  {journal} {Proceedings of the Royal Society of London. Series A.
  Mathematical and Physical Sciences}\ }\textbf {\bibinfo {volume} {281}},\
  \bibinfo {pages} {401} (\bibinfo {year} {1964})}\BibitemShut {NoStop}%
\bibitem [{\citenamefont {Rubtsov}\ \emph {et~al.}(2008)\citenamefont
  {Rubtsov}, \citenamefont {Katsnelson},\ and\ \citenamefont
  {Lichtenstein}}]{Rubtsov08}%
  \BibitemOpen
  \bibfield  {author} {\bibinfo {author} {\bibfnamefont {A.~N.}\ \bibnamefont
  {Rubtsov}}, \bibinfo {author} {\bibfnamefont {M.~I.}\ \bibnamefont
  {Katsnelson}}, \ and\ \bibinfo {author} {\bibfnamefont {A.~I.}\ \bibnamefont
  {Lichtenstein}},\ }\href {\doibase 10.1103/PhysRevB.77.033101} {\bibfield
  {journal} {\bibinfo  {journal} {Phys. Rev. B}\ }\textbf {\bibinfo {volume}
  {77}},\ \bibinfo {pages} {033101} (\bibinfo {year} {2008})}\BibitemShut
  {NoStop}%
\bibitem [{\citenamefont {Toschi}\ \emph {et~al.}(2007)\citenamefont {Toschi},
  \citenamefont {Katanin},\ and\ \citenamefont {Held}}]{Toschi07}%
  \BibitemOpen
  \bibfield  {author} {\bibinfo {author} {\bibfnamefont {A.}~\bibnamefont
  {Toschi}}, \bibinfo {author} {\bibfnamefont {A.~A.}\ \bibnamefont {Katanin}},
  \ and\ \bibinfo {author} {\bibfnamefont {K.}~\bibnamefont {Held}},\ }\href
  {\doibase 10.1103/PhysRevB.75.045118} {\bibfield  {journal} {\bibinfo
  {journal} {Physical Review B (Condensed Matter and Materials Physics)}\
  }\textbf {\bibinfo {volume} {75}},\ \bibinfo {eid} {045118} (\bibinfo {year}
  {2007})}\BibitemShut {NoStop}%
\bibitem [{\citenamefont {Rohringer}\ \emph {et~al.}(2013)\citenamefont
  {Rohringer}, \citenamefont {Toschi}, \citenamefont {Hafermann}, \citenamefont
  {Held}, \citenamefont {Anisimov},\ and\ \citenamefont
  {Katanin}}]{Rohringer13}%
  \BibitemOpen
  \bibfield  {author} {\bibinfo {author} {\bibfnamefont {G.}~\bibnamefont
  {Rohringer}}, \bibinfo {author} {\bibfnamefont {A.}~\bibnamefont {Toschi}},
  \bibinfo {author} {\bibfnamefont {H.}~\bibnamefont {Hafermann}}, \bibinfo
  {author} {\bibfnamefont {K.}~\bibnamefont {Held}}, \bibinfo {author}
  {\bibfnamefont {V.~I.}\ \bibnamefont {Anisimov}}, \ and\ \bibinfo {author}
  {\bibfnamefont {A.~A.}\ \bibnamefont {Katanin}},\ }\href {\doibase
  10.1103/PhysRevB.88.115112} {\bibfield  {journal} {\bibinfo  {journal} {Phys.
  Rev. B}\ }\textbf {\bibinfo {volume} {88}},\ \bibinfo {pages} {115112}
  (\bibinfo {year} {2013})}\BibitemShut {NoStop}%
\bibitem [{\citenamefont {Taranto}\ \emph {et~al.}(2014)\citenamefont
  {Taranto}, \citenamefont {Andergassen}, \citenamefont {Bauer}, \citenamefont
  {Held}, \citenamefont {Katanin}, \citenamefont {Metzner}, \citenamefont
  {Rohringer},\ and\ \citenamefont {Toschi}}]{Taranto14}%
  \BibitemOpen
  \bibfield  {author} {\bibinfo {author} {\bibfnamefont {C.}~\bibnamefont
  {Taranto}}, \bibinfo {author} {\bibfnamefont {S.}~\bibnamefont
  {Andergassen}}, \bibinfo {author} {\bibfnamefont {J.}~\bibnamefont {Bauer}},
  \bibinfo {author} {\bibfnamefont {K.}~\bibnamefont {Held}}, \bibinfo {author}
  {\bibfnamefont {A.}~\bibnamefont {Katanin}}, \bibinfo {author} {\bibfnamefont
  {W.}~\bibnamefont {Metzner}}, \bibinfo {author} {\bibfnamefont
  {G.}~\bibnamefont {Rohringer}}, \ and\ \bibinfo {author} {\bibfnamefont
  {A.}~\bibnamefont {Toschi}},\ }\href {\doibase
  10.1103/PhysRevLett.112.196402} {\bibfield  {journal} {\bibinfo  {journal}
  {Phys. Rev. Lett.}\ }\textbf {\bibinfo {volume} {112}},\ \bibinfo {pages}
  {196402} (\bibinfo {year} {2014})}\BibitemShut {NoStop}%
\bibitem [{\citenamefont {Wentzell}\ \emph {et~al.}(2015)\citenamefont
  {Wentzell}, \citenamefont {Taranto}, \citenamefont {Katanin}, \citenamefont
  {Toschi},\ and\ \citenamefont {Andergassen}}]{Wentzell15}%
  \BibitemOpen
  \bibfield  {author} {\bibinfo {author} {\bibfnamefont {N.}~\bibnamefont
  {Wentzell}}, \bibinfo {author} {\bibfnamefont {C.}~\bibnamefont {Taranto}},
  \bibinfo {author} {\bibfnamefont {A.}~\bibnamefont {Katanin}}, \bibinfo
  {author} {\bibfnamefont {A.}~\bibnamefont {Toschi}}, \ and\ \bibinfo {author}
  {\bibfnamefont {S.}~\bibnamefont {Andergassen}},\ }\href {\doibase
  10.1103/PhysRevB.91.045120} {\bibfield  {journal} {\bibinfo  {journal} {Phys.
  Rev. B}\ }\textbf {\bibinfo {volume} {91}},\ \bibinfo {pages} {045120}
  (\bibinfo {year} {2015})}\BibitemShut {NoStop}%
\bibitem [{\citenamefont {Antipov}\ \emph {et~al.}(2014)\citenamefont
  {Antipov}, \citenamefont {Gull},\ and\ \citenamefont {Kirchner}}]{Antipov14}%
  \BibitemOpen
  \bibfield  {author} {\bibinfo {author} {\bibfnamefont {A.~E.}\ \bibnamefont
  {Antipov}}, \bibinfo {author} {\bibfnamefont {E.}~\bibnamefont {Gull}}, \
  and\ \bibinfo {author} {\bibfnamefont {S.}~\bibnamefont {Kirchner}},\ }\href
  {\doibase 10.1103/PhysRevLett.112.226401} {\bibfield  {journal} {\bibinfo
  {journal} {Phys. Rev. Lett.}\ }\textbf {\bibinfo {volume} {112}},\ \bibinfo
  {pages} {226401} (\bibinfo {year} {2014})}\BibitemShut {NoStop}%
\bibitem [{\citenamefont {Yudin}\ \emph {et~al.}(2014)\citenamefont {Yudin},
  \citenamefont {Hirschmeier}, \citenamefont {Hafermann}, \citenamefont
  {Eriksson}, \citenamefont {Lichtenstein},\ and\ \citenamefont
  {Katsnelson}}]{Yudin14}%
  \BibitemOpen
  \bibfield  {author} {\bibinfo {author} {\bibfnamefont {D.}~\bibnamefont
  {Yudin}}, \bibinfo {author} {\bibfnamefont {D.}~\bibnamefont {Hirschmeier}},
  \bibinfo {author} {\bibfnamefont {H.}~\bibnamefont {Hafermann}}, \bibinfo
  {author} {\bibfnamefont {O.}~\bibnamefont {Eriksson}}, \bibinfo {author}
  {\bibfnamefont {A.~I.}\ \bibnamefont {Lichtenstein}}, \ and\ \bibinfo
  {author} {\bibfnamefont {M.~I.}\ \bibnamefont {Katsnelson}},\ }\href
  {\doibase 10.1103/PhysRevLett.112.070403} {\bibfield  {journal} {\bibinfo
  {journal} {Phys. Rev. Lett.}\ }\textbf {\bibinfo {volume} {112}},\ \bibinfo
  {pages} {070403} (\bibinfo {year} {2014})}\BibitemShut {NoStop}%
\bibitem [{\citenamefont {Si}\ and\ \citenamefont {Smith}(1996)}]{Si96}%
  \BibitemOpen
  \bibfield  {author} {\bibinfo {author} {\bibfnamefont {Q.}~\bibnamefont
  {Si}}\ and\ \bibinfo {author} {\bibfnamefont {J.~L.}\ \bibnamefont {Smith}},\
  }\href {\doibase 10.1103/PhysRevLett.77.3391} {\bibfield  {journal} {\bibinfo
   {journal} {Phys. Rev. Lett.}\ }\textbf {\bibinfo {volume} {77}},\ \bibinfo
  {pages} {3391} (\bibinfo {year} {1996})}\BibitemShut {NoStop}%
\bibitem [{\citenamefont {Smith}\ and\ \citenamefont {Si}(2000)}]{Smith00}%
  \BibitemOpen
  \bibfield  {author} {\bibinfo {author} {\bibfnamefont {J.~L.}\ \bibnamefont
  {Smith}}\ and\ \bibinfo {author} {\bibfnamefont {Q.}~\bibnamefont {Si}},\
  }\href {\doibase 10.1103/PhysRevB.61.5184} {\bibfield  {journal} {\bibinfo
  {journal} {Phys. Rev. B}\ }\textbf {\bibinfo {volume} {61}},\ \bibinfo
  {pages} {5184} (\bibinfo {year} {2000})}\BibitemShut {NoStop}%
\bibitem [{\citenamefont {Chitra}\ and\ \citenamefont
  {Kotliar}(2000)}]{Chitra00}%
  \BibitemOpen
  \bibfield  {author} {\bibinfo {author} {\bibfnamefont {R.}~\bibnamefont
  {Chitra}}\ and\ \bibinfo {author} {\bibfnamefont {G.}~\bibnamefont
  {Kotliar}},\ }\href {\doibase 10.1103/PhysRevLett.84.3678} {\bibfield
  {journal} {\bibinfo  {journal} {Phys. Rev. Lett.}\ }\textbf {\bibinfo
  {volume} {84}},\ \bibinfo {pages} {3678} (\bibinfo {year}
  {2000})}\BibitemShut {NoStop}%
\bibitem [{\citenamefont {Chitra}\ and\ \citenamefont
  {Kotliar}(2001)}]{Chitra01}%
  \BibitemOpen
  \bibfield  {author} {\bibinfo {author} {\bibfnamefont {R.}~\bibnamefont
  {Chitra}}\ and\ \bibinfo {author} {\bibfnamefont {G.}~\bibnamefont
  {Kotliar}},\ }\href {\doibase 10.1103/PhysRevB.63.115110} {\bibfield
  {journal} {\bibinfo  {journal} {Phys. Rev. B}\ }\textbf {\bibinfo {volume}
  {63}},\ \bibinfo {pages} {115110} (\bibinfo {year} {2001})}\BibitemShut
  {NoStop}%
\bibitem [{\citenamefont {Sun}\ and\ \citenamefont {Kotliar}(2002)}]{Sun02}%
  \BibitemOpen
  \bibfield  {author} {\bibinfo {author} {\bibfnamefont {P.}~\bibnamefont
  {Sun}}\ and\ \bibinfo {author} {\bibfnamefont {G.}~\bibnamefont {Kotliar}},\
  }\href {\doibase 10.1103/PhysRevB.66.085120} {\bibfield  {journal} {\bibinfo
  {journal} {Phys. Rev. B}\ }\textbf {\bibinfo {volume} {66}},\ \bibinfo
  {pages} {085120} (\bibinfo {year} {2002})}\BibitemShut {NoStop}%
\bibitem [{\citenamefont {Ayral}\ and\ \citenamefont
  {Parcollet}(2015)}]{Ayral15}%
  \BibitemOpen
  \bibfield  {author} {\bibinfo {author} {\bibfnamefont {T.}~\bibnamefont
  {Ayral}}\ and\ \bibinfo {author} {\bibfnamefont {O.}~\bibnamefont
  {Parcollet}},\ }\href {\doibase 10.1103/PhysRevB.92.115109} {\bibfield
  {journal} {\bibinfo  {journal} {Phys. Rev. B}\ }\textbf {\bibinfo {volume}
  {92}},\ \bibinfo {pages} {115109} (\bibinfo {year} {2015})}\BibitemShut
  {NoStop}%
\bibitem [{\citenamefont {Rubtsov}\ \emph {et~al.}(2012)\citenamefont
  {Rubtsov}, \citenamefont {Katsnelson},\ and\ \citenamefont
  {Lichtenstein}}]{Rubtsov12}%
  \BibitemOpen
  \bibfield  {author} {\bibinfo {author} {\bibfnamefont {A.~N.}\ \bibnamefont
  {Rubtsov}}, \bibinfo {author} {\bibfnamefont {M.~I.}\ \bibnamefont
  {Katsnelson}}, \ and\ \bibinfo {author} {\bibfnamefont {A.~I.}\ \bibnamefont
  {Lichtenstein}},\ }\href {\doibase 10.1016/j.aop.2012.01.002} {\bibfield
  {journal} {\bibinfo  {journal} {Annals of Physics}\ }\textbf {\bibinfo
  {volume} {327}},\ \bibinfo {pages} {1320} (\bibinfo {year}
  {2012})}\BibitemShut {NoStop}%
\bibitem [{\citenamefont {Lindhard}(1954)}]{Lindhard54}%
  \BibitemOpen
  \bibfield  {author} {\bibinfo {author} {\bibfnamefont {J.}~\bibnamefont
  {Lindhard}},\ }\href@noop {} {\bibfield  {journal} {\bibinfo  {journal} {Kgl.
  Danske Videnskab. Selskab Mat.-Fys. Medd.}\ }\textbf {\bibinfo {volume} {28}}
  (\bibinfo {year} {1954})}\BibitemShut {NoStop}%
\bibitem [{\citenamefont {Ehrenreich}\ and\ \citenamefont
  {Cohen}(1959)}]{Ehrenreich59}%
  \BibitemOpen
  \bibfield  {author} {\bibinfo {author} {\bibfnamefont {H.}~\bibnamefont
  {Ehrenreich}}\ and\ \bibinfo {author} {\bibfnamefont {M.~H.}\ \bibnamefont
  {Cohen}},\ }\href {\doibase 10.1103/PhysRev.115.786} {\bibfield  {journal}
  {\bibinfo  {journal} {Phys. Rev.}\ }\textbf {\bibinfo {volume} {115}},\
  \bibinfo {pages} {786} (\bibinfo {year} {1959})}\BibitemShut {NoStop}%
\bibitem [{\citenamefont {Pines}\ and\ \citenamefont
  {Nozi{\`e}res}(1966)}]{Pines66}%
  \BibitemOpen
  \bibfield  {author} {\bibinfo {author} {\bibfnamefont {D.}~\bibnamefont
  {Pines}}\ and\ \bibinfo {author} {\bibfnamefont {P.}~\bibnamefont
  {Nozi{\`e}res}},\ }\href@noop {} {\emph {\bibinfo {title} {The Theory of
  Quantum Liquids: Normal Fermi liquids}}}\ (\bibinfo  {publisher} {W.A.
  Benjamin, Philadelphia},\ \bibinfo {year} {1966})\BibitemShut {NoStop}%
\bibitem [{\citenamefont {Platzman}\ and\ \citenamefont
  {Wolff}(1973)}]{Platzman73}%
  \BibitemOpen
  \bibfield  {author} {\bibinfo {author} {\bibfnamefont {P.~M.}\ \bibnamefont
  {Platzman}}\ and\ \bibinfo {author} {\bibfnamefont {P.~A.}\ \bibnamefont
  {Wolff}},\ }\href@noop {} {\emph {\bibinfo {title} {Waves and Interactions in
  Solid State Plasmas}}},\ Vol.~\bibinfo {volume} {13}\ (\bibinfo  {publisher}
  {Academic Press, New York},\ \bibinfo {year} {1973})\BibitemShut {NoStop}%
\bibitem [{\citenamefont {Baym}\ and\ \citenamefont {Kadanoff}(1961)}]{Baym61}%
  \BibitemOpen
  \bibfield  {author} {\bibinfo {author} {\bibfnamefont {G.}~\bibnamefont
  {Baym}}\ and\ \bibinfo {author} {\bibfnamefont {L.~P.}\ \bibnamefont
  {Kadanoff}},\ }\href {\doibase 10.1103/PhysRev.124.287} {\bibfield  {journal}
  {\bibinfo  {journal} {Phys. Rev.}\ }\textbf {\bibinfo {volume} {124}},\
  \bibinfo {pages} {287} (\bibinfo {year} {1961})}\BibitemShut {NoStop}%
\bibitem [{\citenamefont {Baym}(1962)}]{Baym62}%
  \BibitemOpen
  \bibfield  {author} {\bibinfo {author} {\bibfnamefont {G.}~\bibnamefont
  {Baym}},\ }\href {\doibase 10.1103/PhysRev.127.1391} {\bibfield  {journal}
  {\bibinfo  {journal} {Phys. Rev.}\ }\textbf {\bibinfo {volume} {127}},\
  \bibinfo {pages} {1391} (\bibinfo {year} {1962})}\BibitemShut {NoStop}%
\bibitem [{\citenamefont {Rubtsov}\ \emph {et~al.}(2005)\citenamefont
  {Rubtsov}, \citenamefont {Savkin},\ and\ \citenamefont
  {Lichtenstein}}]{Rubtsov05}%
  \BibitemOpen
  \bibfield  {author} {\bibinfo {author} {\bibfnamefont {A.~N.}\ \bibnamefont
  {Rubtsov}}, \bibinfo {author} {\bibfnamefont {V.~V.}\ \bibnamefont {Savkin}},
  \ and\ \bibinfo {author} {\bibfnamefont {A.~I.}\ \bibnamefont
  {Lichtenstein}},\ }\href {\doibase 10.1103/PhysRevB.72.035122} {\bibfield
  {journal} {\bibinfo  {journal} {Phys. Rev. B}\ }\textbf {\bibinfo {volume}
  {72}},\ \bibinfo {pages} {035122} (\bibinfo {year} {2005})}\BibitemShut
  {NoStop}%
\bibitem [{\citenamefont {Werner}\ \emph {et~al.}(2006)\citenamefont {Werner},
  \citenamefont {Comanac}, \citenamefont {de' Medici}, \citenamefont {Troyer},\
  and\ \citenamefont {Millis}}]{Werner06}%
  \BibitemOpen
  \bibfield  {author} {\bibinfo {author} {\bibfnamefont {P.}~\bibnamefont
  {Werner}}, \bibinfo {author} {\bibfnamefont {A.}~\bibnamefont {Comanac}},
  \bibinfo {author} {\bibfnamefont {L.}~\bibnamefont {de' Medici}}, \bibinfo
  {author} {\bibfnamefont {M.}~\bibnamefont {Troyer}}, \ and\ \bibinfo {author}
  {\bibfnamefont {A.~J.}\ \bibnamefont {Millis}},\ }\href {\doibase
  10.1103/PhysRevLett.97.076405} {\bibfield  {journal} {\bibinfo  {journal}
  {Phys. Rev. Lett.}\ }\textbf {\bibinfo {volume} {97}},\ \bibinfo {pages}
  {076405} (\bibinfo {year} {2006})}\BibitemShut {NoStop}%
\bibitem [{Note1()}]{Note1}%
  \BibitemOpen
  \bibinfo {note} {Higher order diagrams with a local dual Green's function
  connecting local vertices are also zero.}\BibitemShut {Stop}%
\bibitem [{Note2()}]{Note2}%
  \BibitemOpen
  \bibinfo {note} {The self--consistency condition \protect \textup {\hbox
  {\mathsurround \z@ \protect \normalfont (\ignorespaces \ref {eq:sc:db}\unskip
  \@@italiccorr )}} has been used, see the black crosses in Fig. 21 of
  Ref.~\protect \rev@citealpnum {vanLoon14-2}. The effect compared to
  single--shot calculations was small.}\BibitemShut {Stop}%
\bibitem [{\citenamefont {Hafermann}\ \emph {et~al.}(2009)\citenamefont
  {Hafermann}, \citenamefont {Li}, \citenamefont {Rubtsov}, \citenamefont
  {Katsnelson}, \citenamefont {Lichtenstein},\ and\ \citenamefont
  {Monien}}]{Hafermann09}%
  \BibitemOpen
  \bibfield  {author} {\bibinfo {author} {\bibfnamefont {H.}~\bibnamefont
  {Hafermann}}, \bibinfo {author} {\bibfnamefont {G.}~\bibnamefont {Li}},
  \bibinfo {author} {\bibfnamefont {A.~N.}\ \bibnamefont {Rubtsov}}, \bibinfo
  {author} {\bibfnamefont {M.~I.}\ \bibnamefont {Katsnelson}}, \bibinfo
  {author} {\bibfnamefont {A.~I.}\ \bibnamefont {Lichtenstein}}, \ and\
  \bibinfo {author} {\bibfnamefont {H.}~\bibnamefont {Monien}},\ }\href
  {\doibase 10.1103/PhysRevLett.102.206401} {\bibfield  {journal} {\bibinfo
  {journal} {Phys. Rev. Lett.}\ }\textbf {\bibinfo {volume} {102}},\ \bibinfo
  {pages} {206401} (\bibinfo {year} {2009})}\BibitemShut {NoStop}%
\bibitem [{\citenamefont {van Loon}\ \emph
  {et~al.}(2015{\natexlab{b}})\citenamefont {van Loon}, \citenamefont
  {Hafermann}, \citenamefont {Lichtenstein},\ and\ \citenamefont
  {Katsnelson}}]{vanLoon15}%
  \BibitemOpen
  \bibfield  {author} {\bibinfo {author} {\bibfnamefont {E.~G. C.~P.}\
  \bibnamefont {van Loon}}, \bibinfo {author} {\bibfnamefont {H.}~\bibnamefont
  {Hafermann}}, \bibinfo {author} {\bibfnamefont {A.~I.}\ \bibnamefont
  {Lichtenstein}}, \ and\ \bibinfo {author} {\bibfnamefont {M.~I.}\
  \bibnamefont {Katsnelson}},\ }\href {\doibase 10.1103/PhysRevB.92.085106}
  {\bibfield  {journal} {\bibinfo  {journal} {Phys. Rev. B}\ }\textbf {\bibinfo
  {volume} {92}},\ \bibinfo {pages} {085106} (\bibinfo {year}
  {2015}{\natexlab{b}})}\BibitemShut {NoStop}%
\bibitem [{\citenamefont {Katanin}(2013)}]{Katanin13}%
  \BibitemOpen
  \bibfield  {author} {\bibinfo {author} {\bibfnamefont {A.~A.}\ \bibnamefont
  {Katanin}},\ }\href {http://stacks.iop.org/1751-8121/46/i=4/a=045002}
  {\bibfield  {journal} {\bibinfo  {journal} {Journal of Physics A:
  Mathematical and Theoretical}\ }\textbf {\bibinfo {volume} {46}},\ \bibinfo
  {pages} {045002} (\bibinfo {year} {2013})}\BibitemShut {NoStop}%
\bibitem [{\citenamefont {Hafermann}\ \emph {et~al.}(2013)\citenamefont
  {Hafermann}, \citenamefont {Werner},\ and\ \citenamefont
  {Gull}}]{Hafermann13}%
  \BibitemOpen
  \bibfield  {author} {\bibinfo {author} {\bibfnamefont {H.}~\bibnamefont
  {Hafermann}}, \bibinfo {author} {\bibfnamefont {P.}~\bibnamefont {Werner}}, \
  and\ \bibinfo {author} {\bibfnamefont {E.}~\bibnamefont {Gull}},\ }\href
  {\doibase http://dx.doi.org/10.1016/j.cpc.2012.12.013} {\bibfield  {journal}
  {\bibinfo  {journal} {Computer Physics Communications}\ }\textbf {\bibinfo
  {volume} {184}},\ \bibinfo {pages} {1280 } (\bibinfo {year}
  {2013})}\BibitemShut {NoStop}%
\bibitem [{\citenamefont {Hafermann}(2014)}]{Hafermann14}%
  \BibitemOpen
  \bibfield  {author} {\bibinfo {author} {\bibfnamefont {H.}~\bibnamefont
  {Hafermann}},\ }\href {\doibase 10.1103/PhysRevB.89.235128} {\bibfield
  {journal} {\bibinfo  {journal} {Phys. Rev. B}\ }\textbf {\bibinfo {volume}
  {89}},\ \bibinfo {pages} {235128} (\bibinfo {year} {2014})}\BibitemShut
  {NoStop}%
\bibitem [{\citenamefont {Katanin}\ \emph {et~al.}(2009)\citenamefont
  {Katanin}, \citenamefont {Toschi},\ and\ \citenamefont {Held}}]{Katanin09}%
  \BibitemOpen
  \bibfield  {author} {\bibinfo {author} {\bibfnamefont {A.~A.}\ \bibnamefont
  {Katanin}}, \bibinfo {author} {\bibfnamefont {A.}~\bibnamefont {Toschi}}, \
  and\ \bibinfo {author} {\bibfnamefont {K.}~\bibnamefont {Held}},\ }\href
  {\doibase 10.1103/PhysRevB.80.075104} {\bibfield  {journal} {\bibinfo
  {journal} {Phys. Rev. B}\ }\textbf {\bibinfo {volume} {80}},\ \bibinfo
  {pages} {075104} (\bibinfo {year} {2009})}\BibitemShut {NoStop}%
\bibitem [{\citenamefont {Khurana}(1990)}]{Khurana90}%
  \BibitemOpen
  \bibfield  {author} {\bibinfo {author} {\bibfnamefont {A.}~\bibnamefont
  {Khurana}},\ }\href {\doibase 10.1103/PhysRevLett.64.1990} {\bibfield
  {journal} {\bibinfo  {journal} {Phys. Rev. Lett.}\ }\textbf {\bibinfo
  {volume} {64}},\ \bibinfo {pages} {1990} (\bibinfo {year}
  {1990})}\BibitemShut {NoStop}%
\bibitem [{\citenamefont {Otsuki}\ \emph {et~al.}(2014)\citenamefont {Otsuki},
  \citenamefont {Hafermann},\ and\ \citenamefont {Lichtenstein}}]{Otsuki14}%
  \BibitemOpen
  \bibfield  {author} {\bibinfo {author} {\bibfnamefont {J.}~\bibnamefont
  {Otsuki}}, \bibinfo {author} {\bibfnamefont {H.}~\bibnamefont {Hafermann}}, \
  and\ \bibinfo {author} {\bibfnamefont {A.~I.}\ \bibnamefont {Lichtenstein}},\
  }\href {\doibase 10.1103/PhysRevB.90.235132} {\bibfield  {journal} {\bibinfo
  {journal} {Phys. Rev. B}\ }\textbf {\bibinfo {volume} {90}},\ \bibinfo
  {pages} {235132} (\bibinfo {year} {2014})}\BibitemShut {NoStop}%
\bibitem [{\citenamefont {Mielke}(2015)}]{julich2015}%
  \BibitemOpen
  \bibfield  {author} {\bibinfo {author} {\bibfnamefont {A.}~\bibnamefont
  {Mielke}},\ }in\ \href {https://juser.fz-juelich.de/record/205123} {\emph
  {\bibinfo {booktitle} {{M}any-{B}ody {P}hysics: {F}rom {K}ondo to
  {H}ubbard}}},\ \bibinfo {series} {Modeling and Simulation}, Vol.~\bibinfo
  {volume} {5},\ \bibinfo {editor} {edited by\ \bibinfo {editor} {\bibfnamefont
  {E.}~\bibnamefont {Pavarini}}, \bibinfo {editor} {\bibfnamefont
  {E.}~\bibnamefont {Koch}}, \ and\ \bibinfo {editor} {\bibfnamefont
  {P.}~\bibnamefont {Coleman}}}\ (\bibinfo  {publisher} {Forschungszentrum
  Jülich GmbH Zentralbibliothek, Verlag},\ \bibinfo {address} {Jülich},\
  \bibinfo {year} {2015})\BibitemShut {NoStop}%
\bibitem [{\citenamefont {Ayral}\ \emph {et~al.}(2013)\citenamefont {Ayral},
  \citenamefont {Biermann},\ and\ \citenamefont {Werner}}]{Ayral13}%
  \BibitemOpen
  \bibfield  {author} {\bibinfo {author} {\bibfnamefont {T.}~\bibnamefont
  {Ayral}}, \bibinfo {author} {\bibfnamefont {S.}~\bibnamefont {Biermann}}, \
  and\ \bibinfo {author} {\bibfnamefont {P.}~\bibnamefont {Werner}},\ }\href
  {\doibase 10.1103/PhysRevB.87.125149} {\bibfield  {journal} {\bibinfo
  {journal} {Phys. Rev. B}\ }\textbf {\bibinfo {volume} {87}},\ \bibinfo
  {pages} {125149} (\bibinfo {year} {2013})}\BibitemShut {NoStop}%
\bibitem [{\citenamefont {Gubernatis}\ \emph {et~al.}(1985)\citenamefont
  {Gubernatis}, \citenamefont {Scalapino}, \citenamefont {Sugar},\ and\
  \citenamefont {Toussaint}}]{Gubernatis85}%
  \BibitemOpen
  \bibfield  {author} {\bibinfo {author} {\bibfnamefont {J.~E.}\ \bibnamefont
  {Gubernatis}}, \bibinfo {author} {\bibfnamefont {D.~J.}\ \bibnamefont
  {Scalapino}}, \bibinfo {author} {\bibfnamefont {R.~L.}\ \bibnamefont
  {Sugar}}, \ and\ \bibinfo {author} {\bibfnamefont {W.~D.}\ \bibnamefont
  {Toussaint}},\ }\href {\doibase 10.1103/PhysRevB.32.103} {\bibfield
  {journal} {\bibinfo  {journal} {Phys. Rev. B}\ }\textbf {\bibinfo {volume}
  {32}},\ \bibinfo {pages} {103} (\bibinfo {year} {1985})}\BibitemShut
  {NoStop}%
\bibitem [{\citenamefont {Sch\"uler}\ \emph {et~al.}(2013)\citenamefont
  {Sch\"uler}, \citenamefont {R\"osner}, \citenamefont {Wehling}, \citenamefont
  {Lichtenstein},\ and\ \citenamefont {Katsnelson}}]{Schuler13}%
  \BibitemOpen
  \bibfield  {author} {\bibinfo {author} {\bibfnamefont {M.}~\bibnamefont
  {Sch\"uler}}, \bibinfo {author} {\bibfnamefont {M.}~\bibnamefont {R\"osner}},
  \bibinfo {author} {\bibfnamefont {T.~O.}\ \bibnamefont {Wehling}}, \bibinfo
  {author} {\bibfnamefont {A.~I.}\ \bibnamefont {Lichtenstein}}, \ and\
  \bibinfo {author} {\bibfnamefont {M.~I.}\ \bibnamefont {Katsnelson}},\ }\href
  {\doibase 10.1103/PhysRevLett.111.036601} {\bibfield  {journal} {\bibinfo
  {journal} {Phys. Rev. Lett.}\ }\textbf {\bibinfo {volume} {111}},\ \bibinfo
  {pages} {036601} (\bibinfo {year} {2013})}\BibitemShut {NoStop}%
\bibitem [{\citenamefont {Brener}\ \emph {et~al.}(2008)\citenamefont {Brener},
  \citenamefont {Hafermann}, \citenamefont {Rubtsov}, \citenamefont
  {Katsnelson},\ and\ \citenamefont {Lichtenstein}}]{Brener08}%
  \BibitemOpen
  \bibfield  {author} {\bibinfo {author} {\bibfnamefont {S.}~\bibnamefont
  {Brener}}, \bibinfo {author} {\bibfnamefont {H.}~\bibnamefont {Hafermann}},
  \bibinfo {author} {\bibfnamefont {A.~N.}\ \bibnamefont {Rubtsov}}, \bibinfo
  {author} {\bibfnamefont {M.~I.}\ \bibnamefont {Katsnelson}}, \ and\ \bibinfo
  {author} {\bibfnamefont {A.~I.}\ \bibnamefont {Lichtenstein}},\ }\href
  {\doibase 10.1103/PhysRevB.77.195105} {\bibfield  {journal} {\bibinfo
  {journal} {Phys. Rev. B}\ }\textbf {\bibinfo {volume} {77}},\ \bibinfo {eid}
  {195105} (\bibinfo {year} {2008})}\BibitemShut {NoStop}%
\bibitem [{\citenamefont {Hirschmeier}\ \emph {et~al.}(2015)\citenamefont
  {Hirschmeier}, \citenamefont {Hafermann}, \citenamefont {Gull}, \citenamefont
  {Lichtenstein},\ and\ \citenamefont {Antipov}}]{Hirschmeier15}%
  \BibitemOpen
  \bibfield  {author} {\bibinfo {author} {\bibfnamefont {D.}~\bibnamefont
  {Hirschmeier}}, \bibinfo {author} {\bibfnamefont {H.}~\bibnamefont
  {Hafermann}}, \bibinfo {author} {\bibfnamefont {E.}~\bibnamefont {Gull}},
  \bibinfo {author} {\bibfnamefont {A.~I.}\ \bibnamefont {Lichtenstein}}, \
  and\ \bibinfo {author} {\bibfnamefont {A.~E.}\ \bibnamefont {Antipov}},\
  }\href {\doibase 10.1103/PhysRevB.92.144409} {\bibfield  {journal} {\bibinfo
  {journal} {Phys. Rev. B}\ }\textbf {\bibinfo {volume} {92}},\ \bibinfo
  {pages} {144409} (\bibinfo {year} {2015})}\BibitemShut {NoStop}%
\bibitem [{\citenamefont {Bauer}\ \emph {et~al.}(2011)\citenamefont {Bauer},
  \citenamefont {Carr}, \citenamefont {Evertz}, \citenamefont {Feiguin},
  \citenamefont {Freire}, \citenamefont {Fuchs}, \citenamefont {Gamper},
  \citenamefont {Gukelberger}, \citenamefont {Gull}, \citenamefont {Guertler},
  \citenamefont {Hehn}, \citenamefont {Igarashi}, \citenamefont {Isakov},
  \citenamefont {Koop}, \citenamefont {Ma}, \citenamefont {Mates},
  \citenamefont {Matsuo}, \citenamefont {Parcollet}, \citenamefont
  {Pawłowski}, \citenamefont {Picon}, \citenamefont {Pollet}, \citenamefont
  {Santos}, \citenamefont {Scarola}, \citenamefont {Schollwöck}, \citenamefont
  {Silva}, \citenamefont {Surer}, \citenamefont {Todo}, \citenamefont {Trebst},
  \citenamefont {Troyer}, \citenamefont {Wall}, \citenamefont {Werner},\ and\
  \citenamefont {Wessel}}]{ALPS2}%
  \BibitemOpen
  \bibfield  {author} {\bibinfo {author} {\bibfnamefont {B.}~\bibnamefont
  {Bauer}}, \bibinfo {author} {\bibfnamefont {L.~D.}\ \bibnamefont {Carr}},
  \bibinfo {author} {\bibfnamefont {H.~G.}\ \bibnamefont {Evertz}}, \bibinfo
  {author} {\bibfnamefont {A.}~\bibnamefont {Feiguin}}, \bibinfo {author}
  {\bibfnamefont {J.}~\bibnamefont {Freire}}, \bibinfo {author} {\bibfnamefont
  {S.}~\bibnamefont {Fuchs}}, \bibinfo {author} {\bibfnamefont
  {L.}~\bibnamefont {Gamper}}, \bibinfo {author} {\bibfnamefont
  {J.}~\bibnamefont {Gukelberger}}, \bibinfo {author} {\bibfnamefont
  {E.}~\bibnamefont {Gull}}, \bibinfo {author} {\bibfnamefont {S.}~\bibnamefont
  {Guertler}}, \bibinfo {author} {\bibfnamefont {A.}~\bibnamefont {Hehn}},
  \bibinfo {author} {\bibfnamefont {R.}~\bibnamefont {Igarashi}}, \bibinfo
  {author} {\bibfnamefont {S.~V.}\ \bibnamefont {Isakov}}, \bibinfo {author}
  {\bibfnamefont {D.}~\bibnamefont {Koop}}, \bibinfo {author} {\bibfnamefont
  {P.~N.}\ \bibnamefont {Ma}}, \bibinfo {author} {\bibfnamefont
  {P.}~\bibnamefont {Mates}}, \bibinfo {author} {\bibfnamefont
  {H.}~\bibnamefont {Matsuo}}, \bibinfo {author} {\bibfnamefont
  {O.}~\bibnamefont {Parcollet}}, \bibinfo {author} {\bibfnamefont
  {G.}~\bibnamefont {Pawłowski}}, \bibinfo {author} {\bibfnamefont {J.~D.}\
  \bibnamefont {Picon}}, \bibinfo {author} {\bibfnamefont {L.}~\bibnamefont
  {Pollet}}, \bibinfo {author} {\bibfnamefont {E.}~\bibnamefont {Santos}},
  \bibinfo {author} {\bibfnamefont {V.~W.}\ \bibnamefont {Scarola}}, \bibinfo
  {author} {\bibfnamefont {U.}~\bibnamefont {Schollwöck}}, \bibinfo {author}
  {\bibfnamefont {C.}~\bibnamefont {Silva}}, \bibinfo {author} {\bibfnamefont
  {B.}~\bibnamefont {Surer}}, \bibinfo {author} {\bibfnamefont
  {S.}~\bibnamefont {Todo}}, \bibinfo {author} {\bibfnamefont {S.}~\bibnamefont
  {Trebst}}, \bibinfo {author} {\bibfnamefont {M.}~\bibnamefont {Troyer}},
  \bibinfo {author} {\bibfnamefont {M.~L.}\ \bibnamefont {Wall}}, \bibinfo
  {author} {\bibfnamefont {P.}~\bibnamefont {Werner}}, \ and\ \bibinfo {author}
  {\bibfnamefont {S.}~\bibnamefont {Wessel}},\ }\href@noop {} {\bibfield
  {journal} {\bibinfo  {journal} {Journal of Statistical Mechanics: Theory and
  Experiment}\ }\textbf {\bibinfo {volume} {2011}},\ \bibinfo {pages} {P05001}
  (\bibinfo {year} {2011})}\BibitemShut {NoStop}%
\end{thebibliography}%

\end{document}